\documentclass[a4paper,11pt]{article}
\pdfoutput=1 % if your are submitting a pdflatex (i.e. if you have
\usepackage{grffile}
\usepackage{aas_macros,braket}
\usepackage{jcappub,natbib} % for details on the use of the package, please
\DeclareUnicodeCharacter{2212}{-}
\usepackage[T1]{fontenc} % if needed
\usepackage{color}
\usepackage{comment}

\usepackage{mathtools}
\usepackage{amsmath}
\allowdisplaybreaks
\usepackage{float}

\def\be{\begin{equation}}
\def\ee{\end{equation}}
\def\bea{\begin{eqnarray}}
\def\eea{\end{eqnarray}}

\def\ba#1\ea{\begin{align}#1\end{align}}

\def\up{\;\raise1.0pt\hbox{$'$}\hskip-6pt\partial\;}
\def\down{\;\overline{\raise1.0pt\hbox{$'$}\hskip-6pt
		\partial}\;}

\newcommand{\f}{\frac}

\def\apjs{Astrophys.\ J. Supp.}

\definecolor{green2}{cmyk}{1, 0, 1, 0.1}

\title{Primordial tensor bispectra in $\mu$-CMB cross-correlations
}

\author[a]{Giorgio Orlando,}
\author[a]{P. Daniel Meerburg,}
\author[b]{Subodh P. Patil}

\affiliation[a]{Van Swinderen Institute for Particle Physics and Gravity, University of
Groningen, Nijenborgh 4, 9747 AG Groningen, The Netherlands}

\affiliation[b]{Instituut-Lorentz for Theoretical Physics, Leiden University,
2333 CA Leiden, The Netherlands}

\emailAdd{g.orlando@rug.nl}
\emailAdd{p.d.meerburg@rug.nl}
\emailAdd{patil@lorentz.leidenuniv.nl}

\abstract{Cross-correlations between Cosmic Microwave Background (CMB) temperature and polarization anisotropies and $\mu$-spectral distortions have been considered to measure (squeezed) primordial scalar bispectra in a range of scales inaccessible to primary CMB bispectra. In this work we address whether it is possible to constrain tensor non-Gaussianities with these cross-correlations. We find that only primordial tensor bispectra with statistical anisotropies leave distinct signatures, while isotropic tensor bispectra leave either vanishing or highly suppressed signatures. We discuss how the angular dependence of squeezed bispectra in terms of the short and long momenta determine the non-zero cross-correlations. We also discuss how these non-vanishing configurations are affected by the way in which primordial bispectra transform under parity. By employing the so-called BipoSH formalism to capture the observational effects of statistical anisotropies, we make Fisher-forecasts to assess the detection prospects from $\mu T$, $\mu E$ and $\mu B$ cross-correlations. Observing statistical anisotropies in squeezed $\langle \gamma \gamma \gamma\rangle$ and $\langle \gamma \gamma \zeta\rangle$ bispectra is going to be challenging as the imprint of tensor perturbations on $\mu$-distortions is subdominant to scalar perturbations, therefore requiring a large, independent amplification of the effect of tensor perturbations in the $\mu$-epoch. In absence of such a mechanism, statistical anisotropies in squeezed $\langle \zeta \zeta \gamma\rangle$ bispectrum are the most relevant sources of $\mu T$, $\mu E$ and $\mu B$ cross-correlations. In particular, we point out that in anisotropic inflationary models where $\langle \zeta \zeta \zeta \rangle$ leaves potentially observable signatures in $\mu T$ and $\mu E$, the detection prospects of $\langle \zeta \zeta \gamma\rangle$ from $\mu B$ are enhanced.%This is relevant as it provides us with information on the underlying inflationary model that is typically not contained in $\langle \zeta \zeta \zeta \rangle$ alone, including the possibility of parity violation.
}

\begin{document}

\maketitle
\flushbottom

\section{Introduction}
It is well-known that the simplest single field slow-roll models of inflation %\cite{Brout:1977ix,Sato:1980yn,Guth:1980zm,Starobinsky:1980te,Linde:1981mu,Albrecht:1982wi} 
predict a negligible level of non-Gaussianity (NG) in the statistics of both the scalar and tensor primordial perturbations \cite{Maldacena:2002vr,Chen:2010xka}. As a result, a net measurement of a non-Gaussian signal would be critical to falsify the simplest scenario and explore the true particle content and symmetry breaking pattern that characterizes the inflationary epoch. At present, the measurements of the bispectrum of the Cosmic Microwave Background (CMB) temperature and polarization anisotropies made by the Planck satellite have provided the tightest constraints on scalar primordial NGs~\cite{Akrami:2019izv}. However, NGs sourced by primordial gravitational waves are poorly constrained and have only been considered for a handful of models (see e.g.  \cite{Shiraishi:2019yux,DeLuca:2019jzc} and refs. therein). These poor constraints are primarily caused by the fact that primordial gravitational waves are best constrained through CMB $B$ modes and attempts to use $B$ modes in search for NGs (partly) sourced by gravitational waves have not yet been made. Current best constraints are derived from the temperature $T$ and polarization $E$-mode measurements, which are dominated by the Gaussian scalar covariance. 

Despite observations that suggest that primordial perturbations from inflation are almost Gaussian, the lack of a net observation of a NG signal does not allow us to disregard any valid alternative scenario. In fact, the current constraints allow for variety of non-conventional models, such as models with non-attractor phases \cite{Mylova:2018yap,Byrnes:2018txb,Carrilho:2019oqg,Ozsoy:2019lyy,Ozsoy:2019slf,Tasinato:2020vdk}, multi-field models (see e.g. the reviews \cite{Wands:2007bd,Byrnes:2010em}), models with extra (spinning) fields \cite{Bordin:2018pca,Dimastrogiovanni:2018gkl} and extra gauge fields \cite{Barnaby:2010vf,Maleknejad:2011jw,Cook:2013xea,Fujita:2014oba,Dimastrogiovanni:2016fuu,Watanabe:2020ctz}, models with non-Bunch Davies initial states \cite{Holman:2007na,Agarwal:2012mq,Akama:2020jko}, and alternate symmetry breaking patterns~\cite{Endlich:2012pz,Endlich:2013jia,Bartolo:2015qvr,Bartolo:2017szm,Bartolo:2018elp,Ricciardone:2016lym,Ricciardone:2017kre,Mirzagholi:2020irt,Celoria:2020diz,Bartolo:2020gsh,Bordin:2020eui,Cabass:2021iii,Celoria:2021cxq}. Some of these alternative descriptions of the inflationary epoch lead to non-negligible primordial bispectra peaking in the so-called squeezed limit, i.e. in momentum configurations where one of the three momenta is much smaller than the other two, indicating a non-zero correlation between large and small scales. The amplitude of NGs associated to this configuration, $f_{\rm NL}^{\rm loc}$, has already been constrained by the {\it Planck} satellite for pure scalar bispectra as $f_{\rm NL}^{\rm loc, sss} = -0.9 \pm 5.1$ \cite{Akrami:2019izv}. Forthcoming CMB experiments involving $B$ modes aim to constrain also $f_{\rm NL}^{\rm loc, tss} \sim 1$, $f_{\rm NL}^{\rm loc, ttt} \sim 1$ \cite{Shiraishi:2019yux}. However, recently several complementary approaches to test cross-correlations between long and short scales have been proposed. One example is the cross-correlation between CMB temperature and polarization anisotropies and $\mu$- and $y$-spectral distortions (SD) \cite{Pajer:2012vz,Emami:2015xqa,Shiraishi:2015lma,Khatri:2015tla,Ota:2016mqd,Ravenni:2017lgw,Cabass:2018jgj,Remazeilles:2021adt}. Another example is the cross-correlation between the anisotropies in the stochastic gravitational wave background (SGWB) and CMB temperature anisotropies \cite{Iacconi:2020yxn,Adshead:2020bji,Malhotra:2020ket,Dimastrogiovanni:2021mfs}. Squeezed NGs may leave observable imprints also on galaxies \cite{Dalal:2007cu,Matarrese:2008nc,Jeong:2012df,Dimastrogiovanni:2014ina} and the 21-cm emission (see e.g. \cite{Munoz:2015eqa}). All these alternative observational channels aim to provide a measure of (squeezed) NG over a range of scales inaccessible by auto-correlating CMB $T$, $E$ and $B$ modes alone.

In this work we aim to review and extend existing analyses on the cross-correlation between $\mu$-spectral distortions and CMB temperature and polarization anisotropies by considering $B$ modes. By admitting statistical anisotropies in scalar and tensor squeezed bispectra, we compute their effects on the $\langle \mu_{\ell_1}T_{\ell_2} \rangle$, $\langle \mu_{\ell_1} E_{\ell_2} \rangle$, $\langle \mu_{\ell_1} B_{\ell_2}\rangle$ cross-correlations. Statistical anisotropies in squeezed NGs induce statistical anisotropies in these cross-correlations, resulting in non-zero off-diagonal ($\ell_1 \neq \ell_2$) values. We will investigate how the angular dependence of anisotropic squeezed bispectra in terms of the short and long momenta influence the non-zero multipole configurations $\ell_1 \ell_2$. We also discuss how these non-vanishing configurations are affected by the way in which primordial bispectra transform under parity transformation. These provide a new way to test theories admitting violation of statistical isotropy and parity symmetry in the primordial universe. To characterize the observational imprints of these statistical anisotropies, we introduce the so-called BipoSH coefficients \cite{Hajian:2003qq,Souradeep:2003qr,Hajian:2005jh} and make Fisher-forecasts to assess detectability. Besides reproducing previous results on the $\langle \zeta \zeta \zeta \rangle$ bispectrum, we show new results on NGs involving primordial gravitational waves. We find that for almost scale-invariant spectra in the $\mu$-distortion window $k \sim 1-10^6 \, \mbox{Mpc}^{-1}$, in order to detect statistical anisotropies in $\langle \gamma \gamma \gamma\rangle$ and $\langle \gamma \gamma \zeta\rangle$ squeezed bispectra, we need an amplification mechanism that is able to enhance the tensor power spectrum by at least six orders of magnitude with respect to the level constrained at the characteristic scales of the {\it Planck} experiment ($k \lesssim 0.05 \, \mbox{Mpc}^{-1}$). In absence of such a mechanism, we must rely on $\langle \zeta \zeta \gamma \rangle$ bispectrum to constrain the tensor sector. We point out that in models where statistical anisotropies in $\langle \zeta \zeta \zeta \rangle$ bispectrum leave potentially observable signatures in $\langle \mu_{\ell_1}T_{\ell_2} \rangle$ and $\langle \mu_{\ell_1} E_{\ell_2} \rangle$ cross-correlations, the detection prospects of statistical anisotropies in $\langle \zeta \zeta \gamma \rangle$ from $\langle \mu_{\ell_1} B_{\ell_2}\rangle$ are enhanced. This last statement is quite intriguing as it is totally model independent. The forecasts we present are valid assuming cosmic variance limited $T$, $E$ and $B$ modes and assuming PIXIE-like noise levels on the $\mu$ modes. Despite introducing generic statistical anisotropies in squeezed bispectra in terms of spin-weighted spherical harmonics (see eqs. \eqref{eq:ansatzsss}-\eqref{eq:ansatzttt}), the analysis can be repeated for a specific inflationary model through the implementation of publicly available numerical codes. Our results may be relevant in sight of the CMB experiments that are going after the first detection of CMB $B$ modes from tensor perturbations (LiteBIRD \cite{Suzuki:2018cuy,Hazumi:2019lys}, PICO \cite{Hanany:2019lle}) and $\mu$-spectral distortions (PIXIE and its advanced iterations \cite{Kogut:2011xw,Kogut:2019vqh} and possible probe class mission proposals \cite{Delabrouille:2019thj}).

The paper is organized as follows. In sec. \ref{sec:preliminaries} we explain the conventions used to introduce statistical anisotropies in squeezed primordial NGs from inflation. We also briefly review CMB temperature and polarization anisotropies and $\mu$-spectral distortions, providing known results and computational conventions employed. In sec. \ref{sec:comp_non_gaus} we compute the effects of statistical anisotropies in (squeezed) primordial bispectra on the cross-correlations between the CMB temperature and polarization anisotropies and $\mu$-spectral distortions. We comment on the results obtained. In sec. \ref{sec:forecast} we derive Fisher-forecasts on the detectability of the signatures discussed in sec. \ref{sec:comp_non_gaus} by employing the BipoSH formalism. In sec. \ref{sec:observation} we consider various phenomenological and model building aspects of our findings. Finally, in sec. \ref{sec:conclusion} we conclude. Some technical details can be found in the appendix.

\section{Preliminaries} \label{sec:preliminaries}

\subsection{Primordial perturbations from inflation}

Here, we provide the conventions used to describe primordial perturbations from inflation. First, we define the Fourier transform decomposition of scalar and tensor perturbations as
\be
\zeta(\vec x) = \int \frac{d^3 k}{(2 \pi)^3} \,  e^{i \vec k \cdot \vec x} \, \zeta_{\vec k} \, ,
\ee
and
\be
\gamma_{ij}(\vec x) = \int \frac{d^3 k}{(2 \pi)^3} \,  e^{i \vec k \cdot \vec x} \, \sum_{\lambda = R/L} \, \left[\gamma_{\vec k}^\lambda \, \epsilon_{ij}^\lambda(\hat k) \right] \, .
\ee
Here, for the purpose of what follows, we are decomposing tensor perturbations in terms of the chiral polarization basis defined through 
\ba
\epsilon_{ij}^{R, L} &= \frac{1}{\sqrt 2} \left[\epsilon_{ij}^+ \pm i \, \epsilon_{ij}^\times \right] \, ,\\
\gamma^{R, L} &= \frac{1}{\sqrt 2} \left[ \gamma_+ \pm i \, \gamma_\times \right] \, ,
\ea
where $\gamma_{+,\times}$ and $\epsilon_{ij}^{+, \times}$ are the usual linear polarizations of tensor perturbations. 

We remind that, if the tensor wave-vector is written in polar coordinates as 
\be
\hat k = (\sin\theta\cos\phi,\sin\theta\sin\phi,\cos\theta)\, ,
\ee
we can define the linear polarization tensors in terms of two unit vectors perpendicular to $\hat k$ as
\ba
\epsilon_{ij}^{+} &= (u_1)_i (u_1)_j - (u_2)_i (u_2)_j \, ,\\
\epsilon_{ij}^{\times} &= (u_1)_i (u_2)_j + (u_2)_i (u_1)_j \, ,
\ea
where 
\be
u_1 = \left(\sin \phi , - \cos \phi, 0\right)\, ,  \qquad u_2 = \begin{cases} \left(\cos \theta \cos \phi , \cos \theta \sin \phi, - \sin \theta \right) \qquad \mbox{if } \theta < \pi/2 \\
 - \left(\cos \theta \cos \phi , \cos \theta \sin \phi, - \sin \theta \right) \qquad \mbox{if } \theta > \pi/2\, .
\end{cases}
\ee
The chiral polarization basis introduced is normalized such that it satisfies the following identities (see e.g. \cite{Alexander:2004wk})
\begin{align}
\epsilon_{ij}^{L}(\vec{k})\epsilon_{L}^{ij}(\vec{k})&=\epsilon_{ij}^{R}(\vec{k})\epsilon_{R}^{ij}(\vec{k}) = 0 \, , \nonumber\\
\epsilon_{ij}^L(\vec{k})\epsilon_R^{ij}(\vec{k})&=2 \, ,\nonumber\\
\epsilon_{ij}^{R}(-\vec{k})&=\epsilon_{ij}^{L}(\vec{k}) \, ,\nonumber\\
\epsilon^{R *}_{ij}(\vec{k})&=\epsilon^{L}_{ij}(\vec{k}),\nonumber\\
\gamma^{R*}_{\vec k}&= \gamma_{- \vec k}^{L}  \, ,\nonumber\\
k_l\epsilon^{mlj}{\epsilon_j^{(\lambda) i}}(\vec{k})&= -i \alpha_\lambda k \epsilon^{(\lambda) im}(\vec{k}) \, ,\label{eq:circ_identities}
\end{align}
where $\alpha_R=+1$ and $\alpha_L=-1$, and $\epsilon^{mlj}$ denotes the Levi-Civita anti-symmetric symbol.

We define the primordial power spectra as
\ba
\langle \zeta_{\vec k_1} \zeta_{\vec k_2}\rangle &= (2 \pi)^3 \delta^{(3)}(\vec k_1+ \vec k_2) \, P_\zeta(\vec k_1) \, , \\
\langle \gamma_{ij}(\vec k_1) \gamma^{ij}(\vec k_2)\rangle &= (2 \pi)^3 \delta^{(3)}(\vec k_1+ \vec k_2) \, P_t(\vec k_1) \, ,
\ea
where 
\be
\gamma_{ij}(\vec k) = \sum_{\lambda = R/L} \, \left[\gamma_{\vec k}^\lambda \, \epsilon_{ij}^\lambda(\hat k) \right] \, .
\ee
As usual, the isotropic parts of scalar and tensor power spectra from inflation can be expressed as
\be \label{eq:power_inflation}
P_{\zeta}(k) = \frac{2 \pi^2}{k^3}  \mathcal A_s(k) \, ,\qquad P_{t}(k) = \frac{2 \pi^2}{k^3} \mathcal A_t(k) \, ,
\ee
where $A_s(k)$ and $A_t(k)$ are dimensionless amplitudes. Here, we are implicitly assuming invariance under translations during inflation.  Notice that we can also define the polarized-power spectra of tensor perturbations
\ba
\langle  \gamma^R_{\vec k_1} \gamma^{R*}_{\vec k_2}\rangle &= (2 \pi)^3 \delta^{(3)}(\vec k_1+ \vec k_2) P_R(\vec k_1) \, , \\
\langle \gamma^L_{\vec k_1} \gamma^{L*}_{\vec k_2}\rangle &= (2 \pi)^3 \delta^{(3)}(\vec k_1+ \vec k_2) P_L(\vec k_1) \, .
\ea
These power spectra can be used to define the quantity $\chi$
\be
\chi = \frac{P_R - P_L}{P_R + P_L} \, ,
\ee
which is usually referred as chirality of tensor perturbations. This gives the asymmetry between the $R$- and $L$-handed power spectra caused by some parity violation mechanism arising in the primordial universe. Assuming parity is a symmetry of the theory, $P_{R, L}$ are related to $P_t$ by
\be
P_{R, L} = \frac{P_t}{4} \, .
\ee
Finally, we define the primoridal bispectra
\ba \label{eq:def_bispectra}
\langle \zeta_{\vec k_1} \zeta_{\vec k_2} \zeta_{\vec k_3}\rangle &= (2 \pi)^3 \delta^{(3)}(\vec k_1+ \vec k_2+\vec k_3) \, B_{\zeta\zeta\zeta}(\vec k_1,\vec k_2,\vec k_3) \nonumber\\
\langle \zeta_{\vec k_1} \zeta_{\vec k_2} \gamma^{\lambda_3}_{\vec k_3}\rangle &= (2 \pi)^3 \delta^{(3)}(\vec k_1+ \vec k_2+ \vec k_3) \, B^{\lambda_3}_{\zeta\zeta\gamma}(\vec k_1, \vec k_2,\vec k_3) \nonumber\\
\langle \gamma^{\lambda_1}_{\vec k_1} \gamma^{\lambda_2}_{\vec k_2} \zeta_{\vec k_3} \rangle &= (2 \pi)^3 \delta^{(3)}(\vec k_1+ \vec k_2+ \vec k_3) \, B^{\lambda_1 \lambda_2}_{\gamma\gamma\zeta}(\vec k_1,\vec k_2,\vec k_3) \nonumber\\
\langle \gamma^{\lambda_1}_{\vec k_1} \gamma^{\lambda_2}_{\vec k_2} \gamma^{\lambda_3}_{\vec k_3}\rangle &= (2 \pi)^3 \delta^{(3)}(\vec k_1+ \vec k_2+\vec k_3) \, B^{\lambda_1 \lambda_2 \lambda_3}_{\gamma\gamma\gamma}(\vec k_1,\vec k_2,\vec k_3) \, ,
\ea
where we assumed invariance under translations. If we account for the invariance under rotations, the bispectra would depend only by the moduli of the momenta. 

\subsection{Statistical anisotropies in squeezed bispectra} \label{sec:non-gaus}

As will become clearer later on, for an inflationary model to be testable via SD-CMB cross-correlations, it is essential for it to have two main features: non-trivial squeezed bispectra %breaking the single-field consistency relations \footnote{When consistency relations are preserved, the leading-order contribution to the squeezed bispectrum amounts to a gauge artifact and its physical contribution to the anisotropies is suppressed (see, e.g., \cite{Maldacena:2002vr,Creminelli:2004yq}).} 
and scale dependent power spectra and bispectra, so that the amplitude of primordial (scalar and tensor) perturbations can grow at the scales sensitive to spectral distortions. Moreover, here we want to introduce statistical anisotropies in primordial correlators. In fact, as we will see later on, introducing statistical anisotropies will turn out to be crucial for tensor bispectra to leave non-negligible signatures on the observables under consideration.

Instead of relying on a specific model, we adopt a phenomenological approach to introduce statistical anisotropies in the squeezed bispectra of primordial perturbations \footnote{Even if we will not consider a specific inflationary model, we want to point out that the amplitude of squeezed primordial bispectra may be severally constrained by soft theorems in models of inflation with given symmetry patterns. See, e.g., the earliest investigations \cite{Maldacena:2002vr,Tanaka:2011aj,Creminelli:2013cga,Pajer:2013ana}, but also the more recent refs. \cite{Sreenath:2014nca,Sreenath:2014nka,Bordin:2017ozj,Bravo:2017wyw,Finelli:2017fml,Cai:2018dkf,Jazayeri:2019nbi,Bravo:2020hde,Suyama:2021adn}, which investigated soft theorems in more general scenarios. It is not the purpose of this work to have a deep look at this issue, which should be taken in mind when constraining the parameter space of a given inflationary scenario.}. As we are admitting statistical anisotropies, we can allow bispectra to depend on the full three wave-vectors $\vec k_i$ appearing inside the bispectra. However, due to the residual translational symmetry, bispectra can be written in terms of only two independent momenta, which in the case of squeezed bispectra is convenient to take as the long and short momenta $\vec k_l$ and $\vec k_s$. Therefore, our bispectra will depend over the long and short modes wave-numbers $k_l$ and $k_s$, and their directions $\hat k_l$ and $\hat k_s$. In particular, the directional dependence can be expressed through an expansion in terms of spin-weighted spherical harmonics (defined as in eq. \eqref{eq:spin_harm_sph}) that capture all the possible ways in which we can introduce statistical anisotropies. Thus, the leading order contribution to the squeezed limit bispectra can be expressed as \footnote{As said, these parametrizations rely on the fact that in the squeezed limit bispectra may depend on the directions of short and long modes only. The spin-weights of the spherical harmonics corresponding to a given angular dependence $\hat k_i$ reflect the spin of the corresponding field $X_{\vec k_i}$ in the bispectrum. Therefore, a spin-$0$ weight is associated to a long scalar, and a spin-$\pm2$ weight is associated to a long tensor (the precise sign of the weight is determined by the polarization state of the long tensor, $-2$ for R-handed tensors, $+2$ for L-handed tensors). In the squeezed limit, the product of two short scalars or tensors in the form $X_{\vec k_s} X_{-\vec k_s}$ is globally a spin-0 field, yielding to a spin-$0$ weight.} 
\ba 
B_{\zeta \zeta \zeta}(-\vec k_l/2 + \vec k_s, -\vec k_l/2 - \vec k_s, \vec k_l)|_{\vec k_l \rightarrow 0}  &= 4 \pi \, \sum_{L_1, M_1} \sum_{L_2, M_2} Y_{L_1 M_1}(\hat{k}_l) \, Y_{L_2 M_2}(\hat{k}_s)  \nonumber \\ 
& \qquad \qquad \qquad  \times  \, f_{L_1,M_1,L_2,M_2}^{\rm s s s}(k_s, k_l) \, P_{\zeta}(k_l) P_{\zeta}(k_s) \, , \label{eq:ansatzsss} 
\ea
\ba
B^{\lambda_3}_{\zeta \zeta \gamma}(-\vec k_l/2 + \vec k_s, -\vec k_l/2 - \vec k_s, \vec k_l)|_{\vec k_l \rightarrow 0}  &= 4 \pi \, {\xi}_{\lambda_3} \sum_{L_1, M_1} \sum_{L_2, M_2} {}_{\pm2} Y_{L_1 M_1}(\hat{k}_l) \, Y_{L_2 M_2}(\hat{k}_s)  \nonumber \\ 
& \qquad \qquad \qquad  \times \, f_{L_1,M_1,L_2,M_2}^{\rm s s t}(k_s, k_l) \, P_{t}(k_l) P_{\zeta}(k_s) \, , \label{eq:ansatzsst} 
\ea
\ba
B^{\lambda_1\lambda_2}_{\gamma \gamma \zeta}(-\vec k_l/2 + \vec k_s, -\vec k_l/2 - \vec k_s, \vec k_l)|_{\vec k_l \rightarrow 0}  &= 4 \pi \, {\xi}_{\lambda_1\lambda_2} \sum_{L_1, M_1} \sum_{L_2, M_2} Y_{L_1 M_1}(\hat{k}_l)  \, Y_{L_2 M_2}(\hat{k}_s)  \nonumber \\ 
& \qquad \qquad \qquad  \times \, f_{L_1,M_1,L_2,M_2}^{\rm t t s}(k_s, k_l) \,   P_{\zeta}(k_l) P_{t}(k_s) \, , \label{eq:ansatztts} 
\ea
\ba
B^{\lambda_1\lambda_2\lambda_3}_{\gamma \gamma \gamma}(-\vec k_l/2 + \vec k_s, -\vec k_l/2 - \vec k_s, \vec k_l)|_{\vec k_l \rightarrow 0}  &= 4 \pi \, {\xi}_{\lambda_1 \lambda_2 \lambda_3}  \sum_{L_1, M_1} \sum_{L_2, M_2} {}_{\pm 2} Y_{L_1 M_1}(\hat{k}_l) \, Y_{L_2 M_2}(\hat{k}_s) \nonumber \\ 
& \qquad \qquad \qquad \times \, f_{L_1,M_1,L_2,M_2}^{\rm t t t}(k_s, k_l)  \, P_{t}(k_l) P_{t}(k_s) \, , \label{eq:ansatzttt}
\ea
where ${\xi}_{\lambda_3}$, ${\xi}_{\lambda_1 \lambda_2}$, ${\xi}_{\lambda_1 \lambda_2 \lambda_3}$ are polarization coefficients sensitive to the polarization states of tensor perturbations appearing in the cosmological correlators, $f^{xxx}_{L_i, M_i}$ are non-Gaussian amplitudes (which in principle may depend on the short and long momenta $k_s$ and $k_l$) and $P_i(k)$ are the isotropic parts of scalar and tensor power spectra as in eq. \eqref{eq:power_inflation}. Having used spherical harmonics to characterize the directional dependencies, a $4 \pi$ normalization factor has been included. 

While a pure scalar bispectrum is insensitive to parity violation unless $L_1 + L_2 = \mbox{odd}$ (see e.g. \cite{Shiraishi:2016mok}, or apply the parity transformation rule of spherical harmonics, eq. \eqref{eq:parity_Y}), the violation of parity symmetry in bispectra involving tensors can be also introduced through the polarization coefficients. In this case, we would say that these coefficients are parity-even when 
\be \label{eq:Aparityeven}
\xi_{L} = \xi_R \, , \qquad \xi_{LL} = \xi_{RR} \, , \qquad \xi_{LR} = \xi_{RL} \, , \qquad \xi_{LLL} = \xi_{RRR} \, , \qquad \xi_{RRL} = \xi_{LLR} \,  , 
\ee
while parity-odd when they obey
\be \label{eq:Aparityodd}
\xi_{L} = - \xi_R \, , \qquad \xi_{LL} = - \xi_{RR} \, , \qquad \xi_{LR} = - \xi_{RL} \, , \qquad \xi_{LLL} = - \xi_{RRR}  \, , \qquad \xi_{RRL} = - \xi_{LLR} \, .
\ee
Bispectra involving tensors may also experience maximum violation of parity through\footnote{Here, we are assuming primordial gravitational waves with a predominant $L$-handed polarization. Alternatively, one can assume a predominant $R$-handed polarization as well.}
\be \label{eq:Anoparity}
B^{L}_{\zeta \zeta \gamma} \gg  B^{R}_{\zeta \zeta \gamma} \, , \qquad\qquad B^{LL}_{\gamma \gamma \zeta} \gg  B^{RR}_{\gamma \gamma \zeta}, B^{RL}_{\gamma \gamma \gamma } \, , \qquad\qquad B^{LLL}_{\gamma \gamma \gamma} \gg  B^{RRR}_{\gamma \gamma \gamma}, B^{LLR}_{\gamma \gamma \gamma}, B^{RRL}_{\gamma \gamma \gamma} \, .
\ee
Notice, also, that the rotationally invariant case of eqs. \eqref{eq:ansatzsss} and \eqref{eq:ansatztts} is recovered in the limit $L_1 = L_2 = M_1 = M_2 = 0$, while a rationally invariant limit of eqs. \eqref{eq:ansatzsst} and \eqref{eq:ansatzttt} can not be defined.

In the following we assume scalar and tensor dimensionless power spectra to obey the power laws
\be
\mathcal A_{s}(k) = G_s \,\mathcal A_{s}(k_{\rm CMB}) \, \left(\frac{k}{k_{SD}}\right)^{n_s -1} \mathcal\, ,
\ee
and
\be \label{eq:tensor_power_law}
\mathcal A_{t}(k) = G_t \, \mathcal A_{t}(k_{\rm CMB}) \, \left(\frac{k}{k_{SD}}\right)^{n_t} \mathcal \, ,
\ee
where the pivot scales $k_{\rm CMB}$ and $k_{\rm SD}$ label characteristic CMB $T$, $E$ and $B$ modes anisotropies and SD scales, respectively. In this work we choose $k_{\rm CMB} = 0.05 \,\mbox{Mpc}^{-1}$ and $k_{\rm SD} = 1 \,\mbox{Mpc}^{-1}$, but an anologous analysis can be performed for different choices of these characteristic scales. In particular, our choice of $k_{\rm SD}$ here reflects the order of magnitude of the smallest primordial tensor mode that source $\mu$ modes (see fig. \ref{fig:transfer_mu}). Also, we introduced the quantities $G_i$'s defined as
\be
G_s = \frac{\mathcal A_{s}(k_{\rm SD})}{\mathcal A_{s}(k_{\rm CMB})} \, , \qquad G_t = \frac{\mathcal A_{t}(k_{\rm SD})}{\mathcal A_{t}(k_{\rm CMB})} \, .
\ee
Physically, they represent the growth factor of scalar (tensor) perturbations on the characteristic SD scale with respect to the CMB scale. For the scalar and tensor amplitudes at the pivot CMB scale we consider the combined Planck + BICEP2/Keck Array BK15 upper limits~\cite{Akrami:2018odb}
\be
\mathcal A_{s}(k_{\rm CMB}) \simeq 2.1 \times 10^{-9} \, , \qquad \qquad \mathcal A_{t}(k_{\rm CMB}) <  0.056 \, \mathcal A_{s}(k_{\rm CMB})  \, .
\ee
We leave the tilts $n_t$ and $n_s-1$ generic. 

\subsection{Review of CMB anisotropies}

Here, we give a brief overview of the physics of the CMB and how we characterize CMB anisotropies. In general, the CMB fluctuation field includes four different polarization states, the so-called Stokes parameters, which are encoded in a $2\times 2$ density matrix \footnote{When we refer to the Stokes parameters, we take only the relative fluctuations over the respective mean value, i.e. $\Delta_T = (\Delta T - T_0)/T_0$ and so on.}
\ba
\rho_{i j}=\frac{1}{2}\left(\begin{array}{cc}
	\Delta_T+\Delta_Q & \Delta_U-i \Delta_V \\
	\Delta_U+i\Delta_V & \Delta_T-\Delta_Q \\
\end{array}
\right)~,
\label{density matrix}
\ea
where $\Delta_T$, $\Delta_Q$, $\Delta_U$, and $\Delta_V$ are the so-called CMB Stokes parameters (see e.g.  \cite{Kosowsky:1994cy}).

%Unpolarized CMB anisotropies are characterized by $\Delta_Q = \Delta_U = \Delta_V = 0$, and the parameter $\Delta_T$ describes the overall radiation intensity. In literature, the quantity $\Delta_T$ is usually called "temperature fluctuation" or "temperature modes".  The $\Delta_Q$ and $\Delta_U$ Stokes parameters describe the linear polarization of the CMB. In particular, taking two orthogonal $(x,y)$ axes on the polarization plane, the $Q$-mode gives the difference in intensity between CMB photons with polarization vectors along the $x$ and $y$ axes respectively, while the $U$-mode gives the difference in intensity between CMB photons with a polarization vector along axes rotated by 45 degrees with respect to the $x$ and $y$ axes. Finally, the $\Delta_V$ Stokes parameter describes the CMB circular polarization or, better to say, it gives the difference in intensity between the two circular polarization modes of the CMB radiation.
CMB fluctuations (both temperature and polarization) are functions of the position and direction on the sky ${\hat n}$, and they can be expanded on the sphere in terms of a spin-weighted basis \cite{Hu:1997hp}
\be
\Delta_T({\hat{n}})=\sum_{\ell, m}a^{I}_{\ell m}Y_{\ell m}({\hat{n}})~,
\ee
\be
\Delta_V({\hat{n}})=\sum_{\ell, m}a^{V}_{\ell m}Y_{\ell m}({\hat{n}})~,
\ee
\be
\Delta_P^{\pm}({\hat{n}})=(\Delta_Q\pm i \Delta_U)({\hat{n}})=\sum_{\ell, m}a^{\pm 2}_{\ell m}\,\,_{\pm 2\, }\!Y_{\ell m}({\hat{n}})\, , \label{P_def}
\ee
where $_{s \,}\!Y_{\ell m}$ denotes again the spin-weighted spherical harmonics. This decomposition is possible since the $\Delta_T$ and $\Delta_V$ polarization fields turn out to be spin-0 fields on the sphere, while the $(\Delta_Q\pm i \Delta_U)$ combination is a spin $\pm2$ field \cite{Hu:1997hp}. In particular, this last feature implies that $\Delta_Q$ and $\Delta_U$ polarization modes are not invariant under a rotation on the polarization plane (while $\Delta_T$ and $\Delta_V$ modes are). In general, we would prefer a description of the CMB polarization in terms of spin-0 quantities that are invariant under rotations. In order to define these quantities, we need to act on $\Delta_P^{\pm}$ the spin raising and lowering operators $\eth$ and $\bar{\eth}$ (see app. \ref{appen:spin_operators}) as
\be \label{E}
\Delta_{E}({\hat{n}})=-\frac{1}{2}\left[\bar{\eth}^2 \Delta_P^{+}({\hat{n}})+\eth^2 \Delta_P^{-}({\hat{n}})\right]~,
\ee
\be \label{B}
\Delta_{B}({\hat{n}})=\frac{i}{2}\left[\bar{\eth}^2 \Delta_P^{+}({\hat{n}})-\eth^2 \Delta_P^{-}({\hat{n}}) \right]~.
\ee
Here, we have introduced the so-called $E$ and $B$ polarization modes. These modes offer an alternative description of CMB linear polarization which, differently from $Q$ and $U$ modes, is invariant under a rotation on the polarization plane. In the following, we will use the $E, B$ modes to refer to the linear polarization field.

The connection between primordial perturbations from inflation and CMB anisotropies is made through a set of Boltzmann equations (see e.g. \cite{Hu:1997hp,Zaldarriaga:1996xe,Dodelson:2003ft}), which describe the time dependent evolution of CMB polarization modes at linear level and predict the expected amount of each polarization mode today. These equations take care of two main contributions: the Compton scattering between CMB photons and electrons and the gravitational redshift which relates CMB anisotropies to primordial perturbations.

In particular, we can define the so-called spherical harmonic coefficients of each (rotationally invariant) CMB mode on the sky as
\be \label{harmonic_dec}
a^{X}_{\ell m} =\int d^2 {\hat{n}} \, Y_{\ell m}({\hat{n}}) \, \Delta_X({\hat{n}})~,
\ee
where $X= T, E, B, V$.

The coefficients of the unpolarized ($X = T$) and $E, B$-mode polarization ($X = E, B$) anisotropies given by the scalar ($\zeta$) and the tensor perturbations ($\gamma^{R, L}$) from inflation, are expressed, respectively, as \cite{Shiraishi:2010sm,Shiraishi:2010kd}
\begin{align} 
a_{\ell m}^{(s) X} &=
4\pi (-i)^{\ell} \int \frac{d^3 \vec{k}}{(2\pi)^{3}}
{\cal T}_{\ell(s)}^{X}(k) \, Y_{\ell m}^*(\hat{k}) \,\zeta_{\vec{k}}  ~, \label{eq:a_CMB_scalar} \\
%----
a_{\ell m}^{(t) X} &=
4\pi (-i)^{\ell} \int \frac{d^3 \vec{k}}{(2\pi)^{3}}
{\cal T}_{\ell(t)}^{X}(k)  \left[ {}_{-2} Y_{\ell m}^*(\hat{k})  \, \gamma_{\vec{k}}^{R} + \left(-1\right)^x  {}_{+ 2} Y_{\ell m}^*(\hat{k}) \, \gamma_{\vec{k}}^{L}  \right]~,  \label{eq:a_CMB_tensor}
%----- 
\end{align}
where ${\cal T}_{\ell (s)}^{X}(k)$ and ${\cal T}_{\ell (t)}^{X}(k)$ are the scalar and tensor CMB transfer functions, respectively, and $x$ takes $0$ ($1$) for $X= T, E$ ($X = B$). Due to the fact that the conventional physics of the CMB is invariant under parity transformations, usually $a_{\ell m}^{V} = 0$. 

It is clear from the equations just introduced that CMB fluctuations are closely related to initial primordial perturbations, which are set by the inflationary epoch, and thus they are a direct probe of the physics of the early universe. We evaluated CMB transfer functions using the publicly available Boltzmann numerical code CAMB \cite{camb_notes} according to the best-fit Planck 2018 LCDM cosmology ($H_0= 67.32 \, \mbox{km}/\mbox{s} \, \mbox{Mpc}^{-1}$, $\Omega_b h^2 = 0.0224$, $\Omega_c h^2 = 0.120$, $\Omega_k h^2 = 0$, $T_{\rm CMB} = 2.7255 \, \mbox{\rm K}$, $\mathcal A_s(k_*) = 2.1 \times 10^{-9}$ at the pivot scale $k_*= 0.05 \,\mbox{Mpc}^{-1}$,  $n_s = 0.966$, $\tau = 0.0543$ \cite{Aghanim:2018eyx}).

\subsection{Review of $\mu$-type spectral distortions}

Next, we provide a brief review of $\mu$-type SD. Primordial perturbations from inflation on super-horizon scales induce acoustic perturbations in the pre-recombination photon-baryon plasma. When these perturbations finally re-enter the horizon, they start to oscillate, dissipating energy in to the photon-baryon plasma -- a phenomenon known as diffusion damping (also called Silk damping) \cite{Sunyaev:1970eu,daly1991spectral,Barrow:1991,Hu:1994bz}. At very high redshift (redshifts $z \gtrsim 10^6$), Compton (and double Compton) scattering in the photon-baryon plasma is efficient enough to maintain kinetic equilibrium even in presence of heat injection. As a consequence, the photon number density distribution is forced to be the that of a Bose-Einstein fluid at equilibrium with zero chemical potential, i.e. a black body spectrum:
\be
n(\nu) = 1/[e^{x} - 1] \, ,
\ee 
where $x = h \nu/k_B T_\gamma$, with $T_\gamma$ denoting the CMB temperature at a given time, and $h$ and $k_B$ are the Planck and Boltzmann constants, respectively.

At redshifts $5 \times 10^4<z < 10^6$ \footnote{At redshifts smaller than $5 \times 10^4$ also Compton scattering becomes inefficient, yielding to another type of distortions as a result of diffusion damping, the so-called $y$-type distortions, which probe the thermal history during recombination and reionization (see e.g. \cite{Chluba:2019kpb} and refs. therein).}, some of the thermalization processes start to be inefficient, while photons can still maintain an internal thermal equilibrium and conserve the photon number density by means of electron-photon elastic Compton scattering. Thus, as a result of heat injection, the CMB photon number density distribution gets a non-zero chemical potential  
\be
n(\nu)|_\mu = 1/[e^{x + \mu(\nu)} - 1] \, .
\ee 
This chemical potential $\mu$ is what we refer to as $\mu$ distortions of the CMB, or $\mu$ modes. Since the heat in the CMB is caused by perturbations seeded by primordial perturbations during inflation, $\mu$ distortions from acoustic dissipation have a primordial origin, and can therefore be expressed in terms of the primordial power spectra \cite{Hu:1994bz, Chluba:2011hw}. 
In full generality, the expectation value of $\mu$ modes in the CMB monopole due to primordial perturbations can be parametrized as \footnote{See e.g. \cite{Sunyaev:1970eu,daly1991spectral,Hu:1994bz,Chluba:2012we} for more details about the derivation of the scalar SD-transfer function due to the dissipation of scalar perturbations, and \cite{Ota:2014hha,Chluba:2014qia,Kite:2020uix} for the same derivation in the case of tensor perturbations. Here, we limit to give the expression of the transfer functions in units of $c=1$ and using the Mpc as the fundamental unit for lengths, times and energies.} 
\be \label{eq:def_SD}
\langle \mu^{\rm primord}(\vec x) \rangle = \frac{1}{2 \pi^2}\, \int_0^\infty dk \, k^2 \, W_i(k) P_i(k) \, ,
\ee
where $i = {\zeta,t}$ and $W_i(k)$ are the SD-transfer function which for scalar and tensor perturbations can be evaluated analytically as (see e.g. \cite{Chluba:2014qia})
\ba \label{eq:window_scalar}
W_\zeta(k) \approx 1.4 \int_{z_{\mu,y}}^\infty \, dz \, \frac{32 k^2}{45 a H \dot \tau} \, D^2 \, 2\sin^2(k r_s) \, e^{- 2 k^2/k_D^2} \, e^{-(z/z_{dc})^{5/2}}
\ea
and 
\ba \label{eq:window_tensor}
W_t(k) \approx 1.4 \int_{z_{\mu,y}}^\infty \, dz \, \frac{4 a H}{45 \dot \tau} \, \mathcal T_\gamma(k \eta) \, \mathcal T_{\theta}(k/\tau') \, \, e^{- \Gamma \eta} \, e^{-(z/z_{dc})^{5/2}} \,  .
\ea
Here we have introduced several quantities: $z_{\mu,y} \simeq 5 \times 10^4$ is the $\mu$-$y$ distortions transition red-shift, and  $z_{dc} \simeq 2 \times 10^6$ is the red-shift at which thermalization processes are very efficient and $\mu$ modes can not arise. $a$ is the usual scale factor
\be
a = \frac{1}{1+z} \, .
\ee
The quantity $\dot \tau$ is the differential optical depth, given by
\be
\dot \tau = \sigma_T N_e c \simeq 4.4 \times 10^{-21} \, (1+z)^3 \, \mbox{sec}^{-1} = 4.5 \times 10^{-7} \, (1+z)^3 \,\mbox{Mpc}^{-1} \, . 
\ee
$D^2$ is the mode-specific efficiency factor
\be
D^2 = \left[ 1 + 4/15 R_{\nu} \right]^{-2} \, ,
\ee
where $R_ \nu \simeq 0.41$ is the fractional contribution of massless neutrinos to the energy density of relativistic species. The quantity $r_s(z)$, defined as
\be
r_s(z) = \frac{1}{\sqrt 3} \int \frac{dt}{a} = \frac{1}{\sqrt 3} \int_0^z \frac{dz'}{H(z')} 
\ee
is the sound horizon at a given redshift. 

$k_D$ is the damping scale
\be
k_D(z) = 4.0 \times 10^{-6} \, (1 + z)^{3/2} \, \mbox{Mpc}^{-1} \, .
\ee
The function $\mathcal T_\gamma(x)$ denotes the tensor transfer function from inflation and is given by
\be
\mathcal T_\gamma(x) = 2 \left\{ \sum_{n=0}^6 a_n \left[ n j_n(x) - x j_{n+1}(x) \right]  \right\}^2 \, ,
\ee
where $j_n(x)$ denote spherical Bessel functions with the numerical coefficients $a_0 = 1$, $a_2 = 0.243807$, $a_4 = 5.28424 \times 10^{-2}$ and $a_6 = 6.13545 \times 10^{-3}$ (The odd values are vanishing).
\be
e^{- \Gamma \eta} \approx 1 \, ,
\ee
\be
T_\theta(\xi) \approx \frac{1 + 4.48 \xi + 91.0 \xi^2}{1 + 4.64 \xi + 90.2 \xi^2 + 100 \xi^3 + 55.0 \xi^4} \, ,
\ee
where $\xi = k/\tau'$, with $\tau' = a \dot \tau = \dot \tau/(1+z)$. 

For smooth power spectra, we can make the approximation $2 \sin^2(k r_s) \simeq 1$, which is very accurate for nearly scale invariant scalar perturbations. In such a case, the scalar transfer function would simplify considerably into \cite{Chluba:2016aln}
\be \label{eq:scalar_transfer_approx}
W^{\rm approx}_\zeta(k) \approx 2.27 \left[ e^{-2 k^2/k_D^2} \right]_{z_{\mu,y}}^{z_{\rm dc}} \, .
\ee
As it is shown e.g. in \cite{Chluba:2014qia}, eqs. \eqref{eq:window_scalar} and \eqref{eq:window_tensor} are efficient and optimal for forecast purposes. 
\begin{figure}
\centering
 \includegraphics[width=0.7\textwidth]{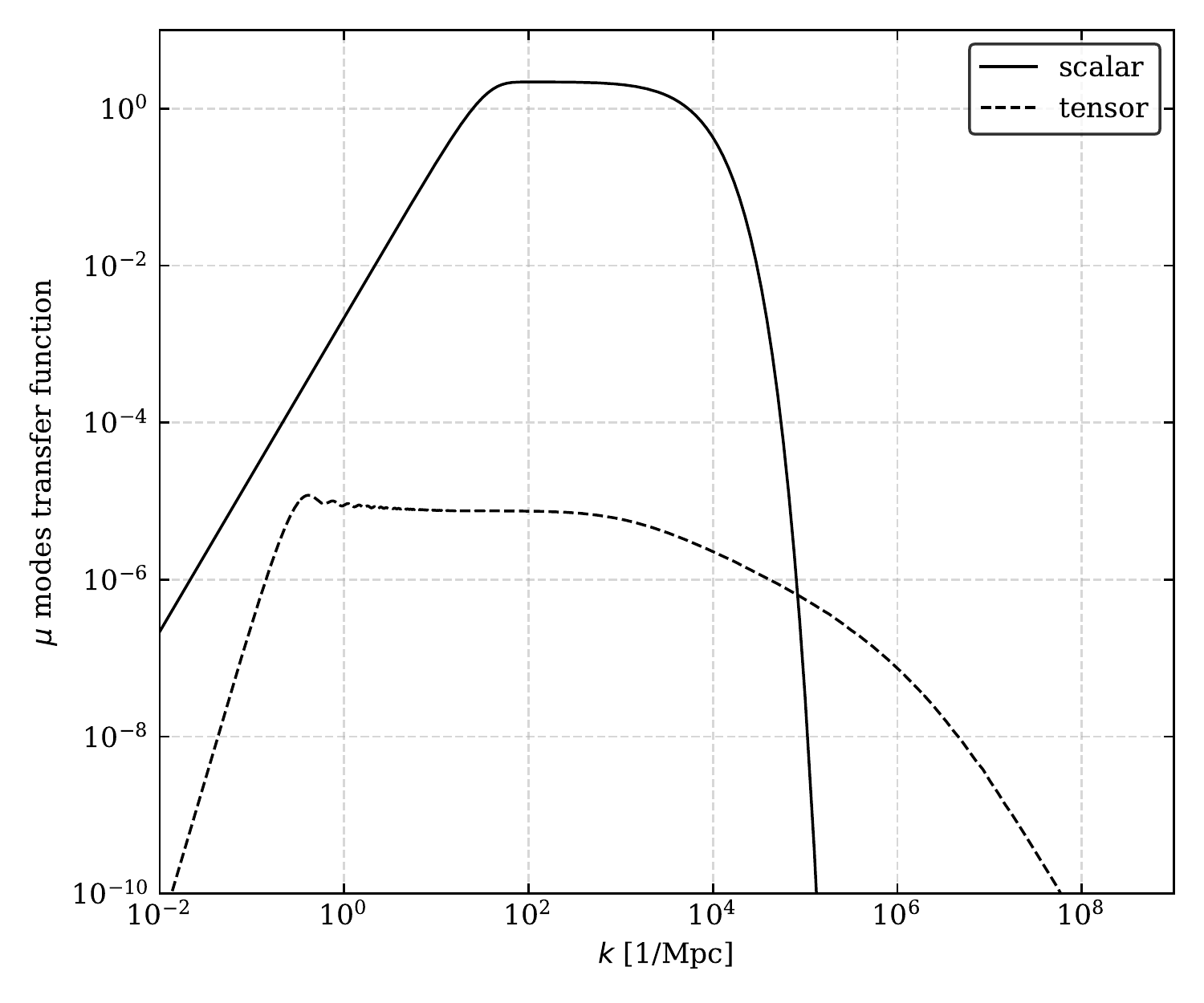}
 \caption{Plot of the $\mu$ modes transfer function for scalar and tensor perturbations.}
\label{fig:transfer_mu}
\end{figure}

Fig. \ref{fig:transfer_mu} shows that tensor perturbations contribute to the generation of $\mu$-distortions over a vast range of scales, $k \simeq 1 - 10^6 \, \mbox{Mpc}^{-1}$, and have a power-law decay for contributions $k > 10^6 \, \mbox{Mpc}^{-1}$. In contrast, the dissipation of scalar perturbations is limited to scales $k \simeq 50 - 10^4 \, \mbox{Mpc}^{-1}$, with a strong exponential decay for contributions $k > 10^4 \, \mbox{Mpc}^{-1}$. The peak of the scalar transfer function is about five orders of magnitude greater than that of the tensor transfer function. These differences in the $\mu$-transfer functions arise because perturbations in the photon fluid sourced by primordial perturbations dissipate their energy differently. In general, perturbations in the photon fluid dissipate through photon-electron scattering and free-streaming effects. However, tensor perturbations dissipate only through free-streaming effects, and so the dissipation cuts off as a power law as rather than an exponential as is the case for scalar modes. Consequently, dissipation of the photon fluid perturbations sourced by tensor perturbations extends over a larger range of scales. Moreover, as transverse, traceless perturbations, tensor perturbations are not significantly attenuated by the CMB photon fluid (as is the case of longitudinal scalar perturbations), hence the tensor dissipation rate is suppressed relative to scalar dissipation. That is, the heat injection is much more inefficient in the case of tensor perturbations, leading to five orders of magnitude difference observed in fig. \ref{fig:transfer_mu}.

By removing the expectation value from eq. \eqref{eq:def_SD}, we can write down its explicit connection to primordial perturbations as
\be \label{eq:mu_scalar}
 \mu^{\rm primord, s}(\vec x) = \int \frac{d^3 \vec{k}_1 \, d^3 \vec{k}_2}{(2\pi)^{6}}
   \, \zeta(\vec{k}_1) \, \zeta(\vec{k}_2) \, \sqrt{W_\zeta\left(k_1\right)} \sqrt{W_\zeta\left(k_2\right)} \, e^{i \vec k_+ \cdot \vec x}
\ee
for scalar perturbations and
\be \label{eq:mu_tensor}
 \mu^{\rm primord, t}(\vec x) = \int \frac{d^3 \vec{k}_1 \, d^3 \vec{k}_2}{(2\pi)^{6}}
   \, \gamma_{ij}(\vec{k}_1) \, \gamma^{ij}(\vec{k}_2) \, \sqrt{W_t\left(k_1\right)} \sqrt{W_t\left(k_2\right)} \, e^{i \vec k_+ \cdot \vec x} 
\ee
for tensor perturbations, where $\vec{k}_+ = \vec{k}_1 + \vec{k}_2$ and $\vec x$ is the position on the last scattering surface. 

As we have done with the CMB $T$, $E$ and $B$ modes, we can make an angular expansion of eqs. \eqref{eq:mu_scalar} and \eqref{eq:mu_tensor} in spherical harmonics as
\be
\label{eq:am_tensor}
a_{\ell m}^{\mu} =  \int d\hat x \, Y^*_{\ell m}(\hat x) \,  \mu^{\rm primord}(\vec x)~,
\ee
which for scalar and tensor perturbations becomes respectively
\be\label{eq:a_mu_scalar}
a_{\ell m}^{\mu, s} = 4 \pi \,(-i)^\ell \, \int \frac{d^3 \vec{k}_1 \, d^3 \vec{k}_2}{(2\pi)^{6}}
   \,  Y^*_{\ell m}(\hat k_+) \, \zeta(\vec{k}_1) \, \zeta(\vec{k}_2) \, \sqrt{W_\zeta\left(k_1\right)} \sqrt{W_\zeta\left(k_2\right)} \, j_\ell(k_+ r_L) 
\ee
and
\be \label{eq:a_mu_tensor}
a_{\ell m}^{\mu, t} = 4 \pi\, (-i)^\ell \, \int \frac{d^3 \vec{k}_1 \, d^3 \vec{k}_2}{(2\pi)^{6}}
   \,  Y^*_{\ell m}(\hat k_+) \, \gamma_{ij}(\vec{k}_1) \, \gamma^{ij}(\vec{k}_2) \, \sqrt{W_t\left(k_1\right)} \sqrt{W_t\left(k_2\right)} \, j_\ell(k_+ r_L) \, .
\ee
Here, $r_L \simeq 1.4 \times 10^4 \, \mbox{Mpc}$ is the distance to the last scattering surface and we have made use of the following identities
\be
e^{i \vec k \cdot \vec x} = \sum_\ell (2 \ell + 1) i^\ell P_\ell(\hat k \cdot \hat x) \, j_\ell(k x) \, ,
\ee
and
\be
P_\ell(\hat k \cdot \hat x) = \frac{4 \pi}{2 \ell + 1} \sum_m \, Y_{\ell m}(\hat k) \, Y^*_{\ell m}(\hat x) \, .
\ee
Using the conventions and the results summarized in this section, we proceed in the next section to compute all possible cross-correlations between CMB $\mu$-spectral distortions (which we henceforth refer to as "SD") and CMB $T$, $E$ and $B$ anisotropies (henceforth referred to as "CMB"). 

\section{Non-Gaussianities from SD-CMB cross-correlations}
\label{sec:comp_non_gaus}
\subsection{Definition of SD-CMB cross-correlations}

We start this section by defining the following $\langle \mu X \rangle$ angular cross-correlation
\ba  \label{eq:defCm}
C^{\mu X}_{\ell_1 \ell_2 m_1 m_2} = \langle a_{\ell_1 m_1}^{\mu} \, a_{\ell_2 m_2}^{X*} \rangle  \, ,
\ea
where $X = T, E, B$.

Note that, by construction, eq. \eqref{eq:defCm} is sensitive to primordial NGs, since it is proportional to the expectation value of the products of three (primordial) fields. In \cite{Pajer:2012vz,Ganc:2012ae,Chluba:2016aln,Dimastrogiovanni:2016aul,Ravenni:2017lgw} the effects of pure scalar NGs in the $\langle \mu T \rangle$ and $\langle \mu E \rangle$ cross-correlations were considered. As emphasized in these references, SD-CMB cross-correlations are sensitive to the squeezed limit of the scalar bispectrum.  
The physical reason behind this is that, even if we cross-correlate CMB anisotropies and $\mu$-distortions at the same angular scales, the primordial perturbations that seeded them refer to very different scales. In particular, $\mu$-distortions are generated by primordial perturbations evaluated at scales much smaller than CMB temperature and polarization anisotropies \footnote{Due to Silk Damping, CMB temperature and polarization anisotropies caused by primordial perturbations are highly suppressed beyond comoving scales $k > 0.15 \, \mbox{Mpc}^{-1}$, whereas observationally significant $\mu$-spectral distortions can be generated by primordial perturbations between comoving scales $k \sim 1-10^6 \, \mbox{Mpc}^{-1}$.}. Schematically, in the cross-correlations of the type \eqref{eq:defCm} $\mu$-distortions and a CMB anisotropies mode $X$ take non-negligible contributions by short and long modes respectively, as
\begin{equation} \label{eq:long_short_scheme}
\mu \propto \zeta_{\vec k_s}\, \zeta_{- \vec k_s}, \,\, \gamma_{\vec k_s} \, \gamma_{- \vec k_s}  \qquad \qquad T, E \propto \zeta_{\vec k_l}, \,\, \gamma_{\vec k_l} \qquad \qquad B \propto  \gamma_{\vec k_l}\, ,
\end{equation}
where we have assumed that CMB $B$ modes are sourced by tensor perturbations only (on large scales).

In the following, we show an original computation of the effects of statistical anisotropies in primordial NGs defined in sec. \ref{sec:non-gaus} on the cross-correlations of the type \eqref{eq:defCm}. In fact the presence of arbitrary breakings of statistical isotropy in our primordial correlators as evident in eqs. \eqref{eq:ansatzsss}-\eqref{eq:ansatzttt} is something new with respect to previous analyses.

\subsection{$\langle \mu T \rangle$}

Here, we focus on the $\langle \mu T \rangle$ (angular) cross-correlation. Mathematically this reads
\ba  \label{eq:defCmT}
C^{\mu T}_{\ell_1 \ell_2 m_1 m_2} = \langle a_{\ell_1 m_1 }^{\mu} \, a_{\ell_2 m_2}^{ T*}\rangle  \, .
\ea
By considering the scheme in eq. \eqref{eq:long_short_scheme} this cross-correlation is affected by all the types of primordial bispectra considered in sec.~\ref{sec:non-gaus}, i.e. schematically
\ba 
C^{\mu T}_{\ell_1 \ell_2 m_1 m_2} \propto \langle \zeta_{\vec k_s} \zeta_{- \vec k_s} \zeta_{\vec k_l} \rangle, \langle \gamma_{\vec k_s} \gamma_{- \vec k_s} \zeta_{\vec k_l} \rangle, \langle \gamma_{\vec k_s} \gamma_{- \vec k_s} \gamma_{\vec k_l} \rangle, \langle \zeta_{\vec k_s} \zeta_{- \vec k_s} \gamma_{\vec k_l} \rangle\, .
\ea
In the following, we compute all these contributions separately.

\vspace{0.5cm}

\centerline{\bf 3-scalars  contribution}

\vspace{0.2cm}

\noindent The contributions from the 3-scalars primordial bispectrum is given by substituting eqs. \eqref{eq:a_CMB_scalar} and \eqref{eq:a_mu_scalar} into \eqref{eq:defCmT}. We get
\ba  \label{eq:CmTsss}
C^{\mu T}_{\ell_1 \ell_2 m_1 m_2} = & i^{\ell_2-\ell_1}  \,16 \pi^2 \int \frac{d^3k \, d^3 k_1 \, d^3 k_2}{(2 \pi)^9} {\cal T}_{\ell_2(s)}^{T}(k) \, j_{\ell_1}(k_+ r_L)  \, \sqrt{W_\zeta\left(k_1\right)} \,\sqrt{W_\zeta\left(k_2\right)}\nonumber\\
& \qquad \qquad \qquad \qquad \qquad \qquad  \times  Y_{\ell_1 m_1}^*(\hat{k}_+) \, Y_{\ell_2 m_2}(\hat{k}) \left[ \braket{\zeta_{\vec{k}_1} \zeta_{\vec{k}_2} \zeta_{- \vec{k}}}  \right] \, .
\ea
We can integrate one of the three momenta by employing the Dirac delta in the definition of primordial bispectra \eqref{eq:def_bispectra}. We obtain
\ba  \label{eq:CmTsss1}
C^{\mu T}_{\ell_1 \ell_2 m_1 m_2} = & i^{\ell_2-\ell_1}  \, 16 \pi^2 \int \frac{d^3 k_1 \, d^3 k_2}{(2 \pi)^6} \, {\cal T}_{\ell_2 (s)}^{T}(k_+) \, j_{\ell_1}(k_+ r_L)  \, \sqrt{W_\zeta\left(k_1\right)} \,\sqrt{W_\zeta\left(k_2\right)}   \nonumber\\
& \qquad \qquad \qquad \qquad \qquad \qquad  \times Y_{\ell_1 m_1}^*(\hat{k}_+) \, Y_{\ell_2 m_2}(\hat{k}_+) \left[ B_{\zeta \zeta \zeta}(\vec k_1, \vec k_2,  - \vec k_+) \right] \, .
\ea
Note that it is much more convenient switching the momenta integrations from $\vec k_1, \vec k_2$ to $\vec k_+, \vec k_-$ by the change of variable $\vec k_\pm = \vec k_1 \pm \vec k_2$. This leads to
\ba  \label{eq:CmTsss2}
C^{\mu T}_{\ell_1 \ell_2 m_1 m_2} = & i^{\ell_2-\ell_1} 2 \pi^2 \int \frac{d^3 k_+ \, d^3 k_-}{(2 \pi)^6} \, {\cal T}_{\ell_2 (s)}^{T}(k_+)  \, j_{\ell_1}(k_+ r_L) \, \sqrt{W_\zeta\left(|(\vec k_+ + \vec k_-)/2|\right)}  \nonumber \\
& \qquad \qquad \qquad \qquad \qquad \times  \,\sqrt{W_\zeta\left(|(\vec k_+ - \vec k_-)/2|\right)} \, Y_{\ell_1 m_1}^*(\hat{k}_+) \,Y_{\ell_2 m_2}( \hat{k}_+) \nonumber\\
&\qquad \qquad \qquad \qquad \qquad \times \left[ B_{\zeta \zeta \zeta}((\vec k_+ + \vec k_-)/2, (\vec k_+ - \vec k_-)/2, - \vec k_+) \right] \, .
\ea
Due to the simultaneous presence of the SD and CMB transfer functions, the integration over $\vec k_+$ and $\vec k_-$ gives a non-negligible contribution only on the very squeezed configurations where $\vec k_+ \rightarrow 0$, i.e. when a scalar wavelength-mode is much greater than the other two scalar modes.

By substituting the squeezed bispectrum \eqref{eq:ansatzsss} into \eqref{eq:CmTsss2}, we get
\ba  \label{eq:ks_zero}
C^{\mu T}_{\ell_1 \ell_2 m_1 m_2} = &  i^{\ell_2-\ell_1} \, 64 \pi^3  \, \int \frac{dk_+ \, d k_-}{(2 \pi)^6} \,k_+^2 \, k_-^2 \, {\cal T}_{\ell_2(s)}^{T}(k_+) \, j_{\ell_1}(k_+ r_L)  \, W_\zeta\left(k_-\right)  P_{\zeta}(k_-) P_{\zeta}(k_+) \nonumber\\
&  \times \, \left[\sum_{L_1, M_1} \sum_{L_2, M_2}  f_{L_1, M_1, L_2, M_2}^{\rm s s s} \, \int d \hat k_- \, Y_{L_2 M_2}(\hat{k}_-) \, \int d \hat k_+  \,  Y_{L_1 M_1}(- \hat{k}_+) \, Y_{\ell_2 m_2}(\hat{k}_+) \,  Y_{\ell_1 m_1}^*(\hat{k}_+)  \right]\, ,
\ea
where we have rescaled the momentum $k_-$ ($\vec k_-' =  \vec k_-/2$). Here, the angular integration over $\hat k_-$ is trivial as it is non-zero only if $L_2 = M_2 = 0$, while the angular integration over $\hat k_+$ can be done in terms of Wigner 3-j symbols (see eq. \eqref{eq:Gaunt_integral} in app. \ref{app:Wigner}) and we get
\ba  
C^{\mu T}_{\ell_1 \ell_2 m_1 m_2} = &  i^{\ell_2-\ell_1}  \, 64 \pi^3 \int \frac{dk_+ \, d k_-}{(2 \pi)^6} \,k_+^2 \, k_-^2 \, {\cal T}_{\ell_2(s)}^{T}(k_+) \, j_{\ell_1}(k_+ r_L)  \, W_\zeta\left(k_-\right)  P_{\zeta}(k_-) P_{\zeta}(k_+) \nonumber \\
& \times \left[\sum_{L_1, M_1}\, f_{L_1,M_1}^{\rm s s s} \, (-1)^{m_1+L_1}  \sqrt{(2 \ell_1 + 1) (2 \ell_2 + 1) (2 L_1 + 1)}  \, \begin{pmatrix}
	\ell_1 & \ell_2 & L_1 \\
	0 & 0 & 0
	\end{pmatrix}\begin{pmatrix}
	\ell_1 & \ell_2 & L_1 \\
	- m_1 & m_2 & M_1
	\end{pmatrix}  \right] \, ,
\ea
where for brevity here and afterwards we will indicate $f_{L_1,M_1}^{x x x} \equiv f_{L_1,M_1, 0, 0}^{x x x}$.

We can rewrite this equation in terms of dimensionless amplitudes as
\ba   \label{eq:mT_sss_final}
C^{\mu T, \,\, \zeta \zeta \zeta}_{\ell_1 \ell_2 m_1 m_2} = &  i^{\ell_2-\ell_1} \, \sum_{L_1, M_1} \, (-1)^{m_1+L_1}  \sqrt{(2 \ell_1 + 1) (2 \ell_2 + 1) (2 L_1 + 1)}  \, \begin{pmatrix}
	\ell_1 & \ell_2 & L_1 \\
	0 & 0 & 0
	\end{pmatrix}\begin{pmatrix}
	\ell_1 & \ell_2 & L_1 \\
	- m_1 & m_2 & M_1
	\end{pmatrix}   \nonumber \\
	& \times \, 4 \pi \, \mathcal I^{\ell_1 \ell_2, L_1 M_1, T}_{\zeta \zeta \zeta} \, ,
\ea
where we defined the integral
\be \label{IT:sss}
\mathcal I^{\ell_1 \ell_2, L_1 M_1, T}_{\zeta \zeta \zeta} = \int d \ln k_+ \, d \ln k_- \,\, {\cal T}_{\ell_2(s)}^{T}(k_+) \, j_{\ell_1}(k_+ r_L)  \, W_\zeta\left(k_-\right)   \, \mathcal A_s(k_-) \, \mathcal A_s(k_+) \, f_{L_1,M_1}^{\rm s s s} \, ,
\ee
which is sensitive to the physical details of a given inflationary model.

\vspace{0.5cm}

\centerline{\bf 1-scalar 2-tensors contribution}

\vspace{0.2cm}

\noindent The contribution from the 1-scalar 2-tensors primordial bispectrum is obtained by substituting eqs. \eqref{eq:a_CMB_scalar} and \eqref{eq:a_mu_tensor} into \eqref{eq:defCmT}. We get
\ba  \label{eq:CmT_tts}
C^{\mu T}_{\ell_1 \ell_2 m_1 m_2} = & i^{\ell_2-\ell_1} \sum_{\lambda, \, \lambda'} \,16 \pi^2 \int \frac{d^3k \, d^3 k_1 \, d^3 k_2}{(2 \pi)^9} {\cal T}_{\ell_2 (s)}^{T}(k) \, j_{\ell_1}(k_+ r_L)  Y_{\ell_1 m_1}^*(\hat{k}_+) \nonumber\\
&\qquad \qquad \qquad \qquad \qquad \qquad \qquad \times \epsilon_{ij}^\lambda(\vec k_1) \epsilon^{ij, \lambda'}(\vec k_2) \, \sqrt{W_t\left(k_1\right)} \,\sqrt{W_t\left(k_2\right)} \nonumber \\
& \qquad \qquad \qquad \qquad \qquad \qquad \qquad  \times  Y_{\ell_2 m_2}(\hat{k}) \left[ \braket{\gamma^\lambda_{\vec{k}_1} \gamma^{\lambda'}_{\vec{k}_2} \zeta_{-\vec{k}}}  \right] \, .
\ea
As before, we can integrate out one of the three momenta, employing the Dirac delta in the definition of the primordial bispectra and switching to the momenta   $\vec k_+, \vec k_-$. We obtain
\ba  \label{eq:CmT_tts3}
C^{\mu T}_{\ell_1 \ell_2 m_1 m_2} = & i^{\ell_2-\ell_1} \sum_{\lambda, \, \lambda'} 2 \pi^2 \int \frac{d^3 k_+ \, d^3 k_-}{(2 \pi)^6} \, {\cal T}_{\ell_2(s)}^{T}(k_+)  \, j_{\ell_1}(k_+ r_L)  \nonumber \\  
& \qquad \qquad \qquad \qquad  \times \sqrt{W_t\left(|(\vec k_+ + \vec k_-)/2|\right)} \sqrt{W_t\left(|(\vec k_+ - \vec k_-)/2|\right)}  \nonumber \\
& \qquad  \qquad \qquad \qquad \times  \epsilon_{ij}^\lambda\left((\vec k_+ + \vec k_-)/2 \right) \, \epsilon^{ij, \lambda'}\left((\vec k_+ - \vec k_-)/2\right) \, Y_{\ell_1 m_1}^*(\hat{k}_+) \,Y_{\ell_2 m_2}(\hat{k}_+) \nonumber\\
&\qquad \qquad \qquad \qquad \qquad \times \left[ B^{\lambda \lambda'}(( \vec k_+ + \vec k_-)/2, ( \vec k_+ - \vec k_-)/2, - \vec k_+) \right] \, .
\ea
Again, the integration over the momenta gives a non-negligible contribution only in the squeezed configurations when the scalar wavelength-mode is much greater than the two tensor modes.

By substituting the squeezed bispectrum \eqref{eq:ansatztts} in eq. \eqref{eq:CmT_tts3}, and expressing the angular integrations in terms of the Wigner 3-j symbols as above, we get the final result
\ba  
C^{\mu T}_{\ell_1 \ell_2 m_1 m_2} = &  i^{\ell_2-\ell_1}  \, 128 \pi^3 \int \frac{dk_+ \, d k_-}{(2 \pi)^6} \,k_+^2 \, k_-^2 \, {\cal T}_{\ell_2(s)}^{T}(k_+) \, j_{\ell_1}(k_+ r_L)  \, W_\zeta\left(k_-\right)  P_{t}(k_-) P_{\zeta}(k_+) \nonumber \\
& \times \left[\sum_{L_1, M_1}\, f_{L_1,M_1}^{\rm s t t} \, (-1)^{m_1+L_1}  \sqrt{(2 \ell_1 + 1) (2 \ell_2 + 1) (2 L_1 + 1)}  \, \begin{pmatrix}
	\ell_1 & \ell_2 & L_1 \\
	0 & 0 & 0
	\end{pmatrix}\begin{pmatrix}
	\ell_1 & \ell_2 & L_1 \\
	- m_1 & m_2 & M_1
	\end{pmatrix}  \right] \nonumber\\
& \times \left(\sum_{\lambda, \, \lambda'}\xi_{\lambda \lambda'}\right) \, .
\ea
We can express this result in terms of dimensionless amplitudes as
\ba   \label{eq:mT_tts_final}
C^{\mu T, \,\, \gamma \gamma \zeta}_{\ell_1 \ell_2 m_1 m_2} = & \sum_{L_1, M_1} \, (-1)^{m_1+L_1}  \sqrt{(2 \ell_1 + 1) (2 \ell_2 + 1) (2 L_1 + 1)}  \, \begin{pmatrix}
	\ell_1 & \ell_2 & L_1 \\
	0 & 0 & 0
	\end{pmatrix}\begin{pmatrix}
	\ell_1 & \ell_2 & L_1 \\
	- m_1 & m_2 & M_1
	\end{pmatrix}  \nonumber \\
& \times \, 8 \pi \, \left(\sum_{\lambda, \, \lambda'}\xi_{\lambda \lambda'}\right)  \,  \mathcal I^{\ell_1 \ell_2, L_1 M_1, T}_{\gamma \gamma \zeta} \, ,
\ea
where
\be \label{IT:tts}
\mathcal I^{\ell_1 \ell_2, L_1 M_1, T}_{\gamma \gamma \zeta} = \int d \ln k_+ \, d \ln k_- \,\, {\cal T}_{\ell_2(s)}^{T}(k_+) \, j_{\ell_1}(k_+ r_L)  \, W_t\left(k_-\right)   \, \mathcal A_t(k_-) \, \mathcal A_s(k_+) \, f_{L_1,M_1}^{\rm s t t} \, .
\ee

\vspace{0.5cm}

\centerline{\bf 2-scalars 1-tensor contribution}

\vspace{0.2cm}

\noindent The contribution from the 2-scalars 1-tensor primordial bispectrum is found by substituting eqs. \eqref{eq:a_CMB_tensor} and \eqref{eq:a_mu_scalar} into \eqref{eq:defCmT}. We get
\ba  \label{eq:CmT_sst1}
C^{\mu T}_{\ell_1 \ell_2 m_1 m_2} = & i^{\ell_2-\ell_1} \, 16 \pi^2 \int \frac{d^3k \, d^3 k_1 \, d^3 k_2}{(2 \pi)^9} {\cal T}_{\ell_2(t)}^{T}(k) \, j_{\ell_1}(k_+ r_L) \, \sqrt{W_\zeta\left(k_1\right)} \,\sqrt{W_\zeta\left(k_2\right)}\nonumber\\
& \qquad\qquad\qquad\quad \times  Y_{\ell_1 m_1}^*(\hat{k}_+) \, \left[{}_{-2} Y_{\ell_2 m_2}(\hat{k}) \braket{\zeta_{\vec{k}_1} \zeta_{\vec{k}_2} \gamma_{-\vec{k}}^{L}} + {}_{+2} Y_{\ell_2 m_2}(\hat{k}) \braket{\zeta_{\vec{k}_1} \zeta_{\vec{k}_2} \gamma_{-\vec{k}}^{R}} \right] \, .
\ea
By integrating out one of the three momenta by employing the Dirac delta in the definition of the primordial bispectra and going through the same steps as above we arrive at
\ba  \label{eq:CmT_sst2}
C^{\mu T}_{\ell_1 \ell_2 m_1 m_2} = & i^{\ell_2-\ell_1} \, 2 \pi^2 \int \frac{d^3 k_+ \, d^3 k_-}{(2 \pi)^6} \, {\cal T}_{\ell_2(t)}^{T}(k_+)  \, j_{\ell_1}(k_+ r_L) \nonumber \\
& \qquad \qquad\qquad \times \sqrt{W_\zeta\left(|(\vec k_+ + \vec k_-)/2|\right)} \,\sqrt{W_\zeta\left(|(\vec k_+ - \vec k_-)/2|\right)} \nonumber \\
& \qquad\qquad\qquad \times Y_{\ell_1 m_1}^*(\hat{k}_+) \, \left[{}_{-2} Y_{\ell_2 m_2}(\hat{k}_+) B^L((\vec k_+ + \vec k_-)/2, (\vec k_+ - \vec k_-)/2,- \vec k_+) \nonumber\right.\\
& \qquad\qquad\qquad\qquad\qquad\qquad \qquad \left. + {}_{+2} Y_{\ell_2 m_2}(\hat{k}_+) B^R((\vec k_+ + \vec k_-)/2, (\vec k_+ - \vec k_-)/2, - \vec k_+) \right] \, .
\ea
By substituting the squeezed bispectrum \eqref{eq:ansatzsst} into \eqref{eq:CmT_sst2}, we can rewrite this as
\ba  
C^{\mu T}_{\ell_1 \ell_2 m_1 m_2} = &  i^{\ell_2-\ell_1} \, 64 \pi^3 \int \frac{dk_+ \, d k_-}{(2 \pi)^6} \,k_+^2 \, k_-^2 \, {\cal T}_{\ell_2(t)}^{T}(k_+) \, j_{\ell_1}(k_+ r_L)  \, W_\zeta\left(k_-\right)  P_{\zeta}(k_-) P_{t}(k_+) \nonumber\\
&  \qquad\qquad \times \sum_{L_1, M_1} \sum_{L_2, M_2}\, f_{L_1, M_1, L_2, M_2}^{\rm s s t}  \int d \hat k_- \, Y_{L_2 M_2}(\hat{k}_-) \nonumber \\
& \qquad\qquad \times \int d \hat k_+  \left[ \xi_L \,\,{}_{+2} Y_{L_1 M_1}(- \hat{k}_+) \, {}_{-2} Y_{\ell_2 m_2}(\hat{k}_+) \,  Y_{\ell_1 m_1}^*(\hat{k}_+) \right. \nonumber\\
&\left. \qquad \qquad\qquad\qquad\qquad \qquad + \xi_R  \,\, {}_{-2} Y_{L_1 M_1}(- \hat{k}_+) \, {}_{+2} Y_{\ell_2 m_2}(\hat{k}_+) \, Y_{\ell_1 m_1}^*(\hat{k}_+) \right] \, .
\ea
As before, we now express the angular integrations in terms of Wigner 3-j symbols
\ba  
C^{\mu T}_{\ell_1 \ell_2 m_1 m_2} = &  i^{\ell_2-\ell_1}  \,  64 \pi^3 \int \frac{dk_+ \, d k_-}{(2 \pi)^6} \,k_+^2 \, k_-^2 \, {\cal T}_{\ell_2(t)}^{T}(k_+) \, j_{\ell_1}(k_+ r_L)  \, W_\zeta\left(k_-\right)  P_{\zeta}(k_-) P_{t}(k_+) \nonumber\\
& \times \, \sum_{L_1, M_1} \, (-1)^{m_1 + L_1} \, f_{L_1, M_1}^{\rm s s t}  \sqrt{(2 \ell_1 + 1) (2 \ell_2 + 1) (2 L_1 + 1)} \, \begin{pmatrix}
	\ell_1 & \ell_2 & L_1 \\
	0 & 2 & -2
	\end{pmatrix}\begin{pmatrix}
	\ell_1 & \ell_2 & L_1  \\
	-m_1 & m_2 & M_1
	\end{pmatrix}  \nonumber\\
&  \times \left[ \xi_L + (-1)^{\ell_1 + \ell_2+L_1} \, \xi_R \right]   \, .
\ea
We can express the final result in terms of dimensionless amplitudes as
\ba   \label{eq:mT_sst_final}
C^{\mu T, \,\, \zeta \zeta \gamma}_{\ell_1 \ell_2 m_1 m_2} = & i^{\ell_2-\ell_1}  \, \sum_{L_1, M_1} \, (-1)^{m_1 + L_1} \,  \sqrt{(2 \ell_1 + 1) (2 \ell_2 + 1) (2 L_1 + 1)} \, \begin{pmatrix}
	\ell_1 & \ell_2 & L_1 \\
	0 & 2 & -2
	\end{pmatrix}\begin{pmatrix}
	\ell_1 & \ell_2 & L_1  \\
	-m_1 & m_2 & M_1
	\end{pmatrix} \nonumber \\
& \times \, 4 \pi \, \mathcal I^{\ell_1 \ell_2, L_1 M_1, T}_{\zeta \zeta \gamma} \, , 	
\ea
where
\ba \label{IT:sst}
\mathcal I^{\ell_1 \ell_2, L_1 M_1, T}_{\zeta \zeta \gamma} = \int d \ln k_+ \, dì \ln k_- \, &{\cal T}_{\ell_2(t)}^{T}(k_+) \, j_{\ell_1}(k_+ r_L)  \, W_\zeta\left(k_-\right)   \, \mathcal A_s(k_-) \, \mathcal A_t(k_+) \, f_{L_1, M_1}^{\rm s s t}  \nonumber \\
& \times \left[ \xi_L + (-1)^{\ell_1 + \ell_2 + L_1} \, \xi_R \right] \, .
\ea

\vspace{0.5cm}

\centerline{\bf 3-tensors contribution}

\vspace{0.2cm}
\noindent The contribution from the 3-tensors primordial bispectrum is given by substituting eqs. \eqref{eq:a_CMB_tensor} and \eqref{eq:a_mu_tensor} into \eqref{eq:defCmT}. We get
\ba  \label{eq:CmTttt1}
C^{\mu T}_{\ell_1 \ell_2 m_1 m_2} = & i^{\ell_2-\ell_1} \,\sum_{\lambda, \, \lambda'}\, 16 \pi^2 \int \frac{d^3k \, d^3 k_1 \, d^3 k_2}{(2 \pi)^9} {\cal T}_{\ell_2(t)}^{T}(k) \, j_{\ell_1}(k_+ r_L)  \,  Y_{\ell_1 m_1}^*(\hat{k}_+) \nonumber \\
&  \qquad\qquad\qquad\qquad\qquad \times \, \epsilon_{ij}^\lambda(\vec k_1) \epsilon^{ij, \lambda'}(\vec k_2) \, \sqrt{W_t\left(k_1\right)} \sqrt{W_t\left(k_2\right)} \, \nonumber\\
& \qquad\qquad\qquad\qquad\qquad \times \left[{}_{-2} Y_{\ell_2 m_2}(\hat{k}) \braket{\mathcal \gamma^\lambda_{\vec{k_1}} \mathcal \gamma^{\lambda'}_{\vec{k_2}} \gamma_{-\vec{k}}^{L}} + {}_{+2} Y_{\ell_2 m_2}(\hat{k}) \braket{\mathcal \gamma^\lambda_{\vec{k_1}} \mathcal \gamma^{\lambda'}_{\vec{k_2}} \gamma_{-\vec{k}}^{R}} \right] \, .
\ea
By going through steps analogous to above we arrive at 
\ba  \label{eq:CmTttt3}
C^{\mu T}_{\ell_1 \ell_2 m_1 m_2} = & i^{\ell_2-\ell_1} \,\sum_{\lambda,\, \lambda'} \, 2 \pi^2 \int \frac{d^3 k_+ \, d^3 k_-}{(2 \pi)^6} \, {\cal T}_{\ell_2(t)}^{T}(k_+)  \, j_{\ell_1}(k_+ r_L) \, \epsilon_{ij}^\lambda\left((\vec k_+ + \vec k_-)/2\right) \nonumber \\
& \qquad\qquad\qquad \times\, \epsilon^{ij, \lambda'}\left((\vec k_+ - \vec k_-)/2\right) \sqrt{W_t\left(|(\vec k_+ + \vec k_-)/2|\right)} \sqrt{W_t\left(|(\vec k_+ - \vec k_-)/2|\right)} \nonumber\\
& \qquad\qquad\qquad \times Y_{\ell_1 m_1}^*(\hat{k}_+) \, \left[{}_{-2} Y_{\ell_2 m_2}(\hat{k}_+) B^{\lambda \lambda' L}((\vec k_+ + \vec k_-)/2, (\vec k_+ - \vec k_-)/2, -\vec k_+) \nonumber\right.\\
& \qquad\qquad\qquad\qquad\qquad\qquad \left. + {}_{+2} Y_{\ell_2 m_2}(\hat{k}_+) B^{\lambda \lambda' R}((\vec k_+ + \vec k_-)/2, (\vec k_+ - \vec k_-)/2, -\vec k_+) \right] \, .
\ea
By substituting the squeezed bispectrum \eqref{eq:ansatzttt} into \eqref{eq:CmTttt3}, we find
\ba  \label{eq:CmTttt4}
C^{\mu T}_{\ell_1 \ell_2 m_1 m_2} = &  i^{\ell_2-\ell_1} \, 128  \pi^3 \int \frac{dk_+ \, d k_-}{(2 \pi)^6} \,k_+^2 \, k_-^2 \, {\cal T}_{\ell_2(t)}^{T}(k_+) \, W_t\left(k_-\right) \, j_{\ell_1}(k_+ r_L)  \,  P_{t}(k_+) P_{t}(k_-) \nonumber\\
& \qquad\qquad \times \, \sum_{L_1, M_1} \sum_{L_2, M_2} \, f_{L_1, M_1, L_2, M_2}^{\rm s s t}  \, \int d \hat k_- \, Y_{L_2 M_2}(\hat{k}_-) \nonumber \\
&\qquad \qquad \times \, \int d \hat k_+ \, \left[(\xi_{LLL} + \xi_{LLR}) \,{}_{+2} Y_{L_1 M_1}(- \hat{k}_+)\, {}_{-2} Y_{\ell_2 m_2}(\hat{k}_+) \,  Y_{\ell_1 m_1}^*(\hat{k}_+) \right. \nonumber\\
&\left. \qquad\qquad\qquad\qquad\qquad\qquad  + (\xi_{RRL} + \xi_{RRR}) \,{}_{-2} Y_{L_1 M_1}(- \hat{k}_+)\, {}_{+2} Y_{\ell_2 m_2}(\hat{k}_+) \,  Y_{\ell_1 m_1}^*(\hat{k}_+) \right] \, . 
\ea
The angular integrations are expressed in terms of Wigner 3-j symbols as above and we obtain the final result
\ba  \label{eq:CmT_ttt_final}
C^{\mu T, \,\, \gamma\gamma\gamma}_{\ell_1 \ell_2 m_1 m_2} = & i^{\ell_2-\ell_1}  \, \sum_{L_1, M_1} \, (-1)^{m_1 + L_1} \,  \sqrt{(2 \ell_1 + 1) (2 \ell_2 + 1) (2 L_1 + 1)} \, \begin{pmatrix}
	\ell_1 & \ell_2 & L_1 \\
	0 & 2 & -2
	\end{pmatrix}\begin{pmatrix}
	\ell_1 & \ell_2 & L_1  \\
	-m_1 & m_2 & M_1
	\end{pmatrix} \nonumber \\
& \times \, 8 \pi \, \mathcal I^{\ell_1 \ell_2, L_1 M_1, T}_{\gamma\gamma\gamma} \, , 
\ea
where
\ba \label{IT:ttt}
\mathcal I^{\ell_1 \ell_2, L_1 M_1, T}_{\gamma \gamma \gamma} = \int d \ln k_+ \, d \ln k_- \, & {\cal T}_{\ell_2(t)}^{T}(k_+) \, j_{\ell_1}(k_+ r_L)  \, W_t\left(k_-\right) \, \mathcal A_t(k_+) \, \mathcal A_t(k_-) \, f_{L_1, M_1}^{\rm t t t}  \nonumber \\
&\times \left[ (\xi_{LLL} + \xi_{LLR}) + (-1)^{\ell_1 + \ell_2 + L_1} \,(\xi_{RRL} + \xi_{RRR}) \right] \, .
\ea

\subsection{$\langle \mu E \rangle$}

Here, we focus on the $\langle \mu E \rangle$ (angular) cross-correlation defined as
\ba  \label{eq:defCmE}
C^{\mu E}_{\ell_1 \ell_2 m_1 m_2} = \langle a_{\ell_1 m_1}^{\mu} \, a_{\ell_2 m_2}^{ E*}\rangle \, .
\ea
The computations resemble the $\langle \mu T \rangle$ case, apart for the substitution $T \rightarrow E$. Therefore, in the following we show only the final results.

\vspace{0.5cm}

\centerline{\bf 3-scalars  contribution}

\vspace{0.2cm}

\ba   \label{eq:mE_sss_final}
C^{\mu E, \,\, \zeta \zeta \zeta}_{\ell_1 \ell_2 m_1 m_2} = & \sum_{L_1, M_1} \, (-1)^{m_1+L_1}  \sqrt{(2 \ell_1 + 1) (2 \ell_2 + 1) (2 L_1 + 1)}  \, \begin{pmatrix}
	\ell_1 & \ell_2 & L_1 \\
	0 & 0 & 0
	\end{pmatrix}\begin{pmatrix}
	\ell_1 & \ell_2 & L_1 \\
	- m_1 & m_2 & M_1
	\end{pmatrix}  \nonumber \\
& \times \, 4 \pi \, \mathcal I^{\ell_1 \ell_2, L_1 M_1, E}_{\zeta\zeta \zeta} \, ,
\ea
where
\be \label{IE:sss}
\mathcal I^{\ell_1 \ell_2, L_1 M_1, E}_{\zeta \zeta \zeta} = \int d \ln k_+ \, d \ln k_- \,\, {\cal T}_{\ell_2(s)}^{E}(k_+) \, j_{\ell_1}(k_+ r_L)  \, W_s\left(k_-\right)   \, \mathcal A_s(k_-) \, \mathcal A_s(k_+) \, f_{L_1,M_1}^{\rm s s s} \, .
\ee

\vspace{0.5cm}

\centerline{\bf 1-scalar 2-tensors contribution}

\vspace{0.2cm}

\ba   \label{eq:mE_tts_final}
C^{\mu E, \,\, \gamma \gamma \zeta}_{\ell_1 \ell_2 m_1 m_2} = & \sum_{L_1, M_1} \, (-1)^{m_1+L_1}  \sqrt{(2 \ell_1 + 1) (2 \ell_2 + 1) (2 L_1 + 1)}  \, \begin{pmatrix}
	\ell_1 & \ell_2 & L_1 \\
	0 & 0 & 0
	\end{pmatrix}\begin{pmatrix}
	\ell_1 & \ell_2 & L_1 \\
	- m_1 & m_2 & M_1
	\end{pmatrix}  \nonumber \\
& \times \, 8 \pi \,\left(\sum_{\lambda, \, \lambda'}\xi_{\lambda \lambda'}\right)  \,  \mathcal I^{\ell_1 \ell_2, L_1 M_1, E}_{\gamma \gamma \zeta} \, ,
\ea
where
\be \label{IE:tts}
\mathcal I^{\ell_1 \ell_2, L_1 M_1, E}_{\gamma \gamma \zeta} = \int d \ln k_+ \, d \ln k_- \,\, {\cal T}_{\ell_2(s)}^{E}(k_+) \, j_{\ell_1}(k_+ r_L)  \, W_t\left(k_-\right)   \, \mathcal A_t(k_-) \, \mathcal A_s(k_+) \, f_{L_1,M_1}^{\rm s t t} \, .
\ee

\vspace{0.5cm}

\centerline{\bf 2-scalars 1-tensor contribution}

\vspace{0.2cm}

\ba  \label{eq:mE_sst_final}
C^{\mu E, \,\, \zeta \zeta \gamma}_{\ell_1 \ell_2 m_1 m_2} = & i^{\ell_2-\ell_1}  \, \sum_{L_1, M_1} \, (-1)^{m_1 + L_1} \,  \sqrt{(2 \ell_1 + 1) (2 \ell_2 + 1) (2 L_1 + 1)} \, \begin{pmatrix}
	\ell_1 & \ell_2 & L_1 \\
	0 & 2 & -2
	\end{pmatrix}\begin{pmatrix}
	\ell_1 & \ell_2 & L_1  \\
	-m_1 & m_2 & M_1
	\end{pmatrix} \nonumber \\
& \times \, 4 \pi \, \mathcal I^{\ell_1 \ell_2, L_1 M_1, E}_{\zeta \zeta \gamma} \, , 
\ea
where
\ba \label{IE:sst}
\mathcal I^{\ell_1 \ell_2, L_1 M_1, E}_{\zeta \zeta \gamma} = \int d \ln k_+ \, d \ln k_- \, & {\cal T}_{\ell_2(t)}^{E}(k_+) \, j_{\ell_1}(k_+ r_L)  \, W_t\left(k_-\right) \, \mathcal A_t(k_+) \, \mathcal A_s(k_-) \, f_{L_1, M_1}^{\rm s s t}  \nonumber \\
&\times \left[ \xi_L + (-1)^{\ell_1 + \ell_2 + L_1} \, \xi_R \right]\, .
\ea

\vspace{0.5cm}

\centerline{\bf 3-tensors contribution}

\vspace{0.2cm}

\ba  \label{eq:CmE_ttt_final}
C^{\mu E, \,\, \gamma\gamma\gamma}_{\ell_1 \ell_2 m_1 m_2} = & i^{\ell_2-\ell_1}  \, \sum_{L_1, M_1} \, (-1)^{m_1 + L_1} \,  \sqrt{(2 \ell_1 + 1) (2 \ell_2 + 1) (2 L_1 + 1)} \, \begin{pmatrix}
	\ell_1 & \ell_2 & L_1 \\
	0 & 2 & -2
	\end{pmatrix}\begin{pmatrix}
	\ell_1 & \ell_2 & L_1  \\
	-m_1 & m_2 & M_1
	\end{pmatrix} \nonumber \\
& \times \, 8 \pi \, \mathcal I^{\ell_1 \ell_2, L_1 M_1, E}_{\gamma\gamma\gamma} \, , 
\ea
where
\ba \label{IE:ttt}
\mathcal I^{\ell_1 \ell_2, L_1 M_1, E}_{\gamma \gamma \gamma} = \int d \ln k_+ \, d \ln k_- \, & {\cal T}_{\ell_2(t)}^{E}(k_+) \, j_{\ell_1}(k_+ r_L)  \, W_t\left(k_-\right) \, \mathcal A_t(k_+) \, \mathcal A_t(k_-) \, f_{L_1, M_1}^{\rm t t t}  \nonumber \\
&\times \left[ (\xi_{LLL} + \xi_{LLR}) + (-1)^{\ell_1 + \ell_2 + L_1} \,(\xi_{RRL} + \xi_{RRR}) \right] \, .
\ea

\subsection{$\langle \mu B \rangle$}

Here, we focus on the $\langle \mu B \rangle$ (angular) cross-correlation defined as
\ba  \label{eq:defCmB}
C^{\mu B}_{\ell_1 \ell_2 m_1 m_2} = \langle a_{\ell_1 m_1}^{\mu} \, a_{\ell_2 m_2}^{(t) B*}\rangle  \, .
\ea
Again, following the scheme \eqref{eq:long_short_scheme}, this cross-correlation will be proportional to the 1-graviton 2 scalars and 3-gravitons squeezed bispectra
\ba 
C^{\mu B}_{\ell_1 \ell_2 m_1 m_2} \propto \langle \zeta_{\vec k_s} \zeta_{- \vec k_s} \gamma_{\vec k_l} \rangle, \langle \gamma_{\vec k_s} \gamma_{- \vec k_s} \gamma_{\vec k_l} \rangle \, .
\ea
The computations resemble the $\langle \mu T \rangle$ case, apart for the exchange $T \rightarrow B$ and the sign flip $+ \rightarrow -$ inside the square parenthesis in eqs. \eqref{IT:sst} and \eqref{IT:ttt}. Therefore, in the following we give only the final results.

\vspace{0.5cm}

\centerline{\bf 2-scalars 1-tensor contribution}

\vspace{0.2cm}

\ba  \label{eq:mB_sst_final}
C^{\mu B, \,\, \zeta \zeta \gamma}_{\ell_1 \ell_2 m_1 m_2} = & i^{\ell_2-\ell_1}  \, \sum_{L_1, M_1} \, (-1)^{m_1 + L_1} \,  \sqrt{(2 \ell_1 + 1) (2 \ell_2 + 1) (2 L_1 + 1)} \, \begin{pmatrix}
	\ell_1 & \ell_2 & L_1 \\
	0 & 2 & -2
	\end{pmatrix}\begin{pmatrix}
	\ell_1 & \ell_2 & L_1  \\
	-m_1 & m_2 & M_1
	\end{pmatrix} \nonumber \\
& \times \, 4 \pi \, \mathcal I^{\ell_1 \ell_2, L_1 M_1, B}_{\zeta \zeta \gamma} \, , 
\ea
where
\ba \label{IB:sst}
\mathcal I^{\ell_1 \ell_2, L_1 M_1, B}_{\zeta \zeta \gamma} = \int d \ln k_+ \, d \ln k_- \, & {\cal T}_{\ell_2(t)}^{B}(k_+) \, j_{\ell_1}(k_+ r_L)  \, W_t\left(k_-\right) \, \mathcal A_t(k_+) \, \mathcal A_s(k_-) \, f_{L_1, M_1}^{\rm s s t}  \nonumber \\
&\times \left[ \xi_L - (-1)^{\ell_1 + \ell_2 + L_1} \, \xi_R \right]\, .
\ea

\vspace{0.5cm}

\centerline{\bf 3-tensors contribution}

\vspace{0.2cm}

\ba  \label{eq:CmB_ttt_final}
C^{\mu B, \,\, \gamma\gamma\gamma}_{\ell_1 \ell_2 m_1 m_2} = & i^{\ell_2-\ell_1}  \, \sum_{L_1, M_1} \, (-1)^{m_1 + L_1} \,  \sqrt{(2 \ell_1 + 1) (2 \ell_2 + 1) (2 L_1 + 1)} \, \begin{pmatrix}
	\ell_1 & \ell_2 & L_1 \\
	0 & 2 & -2
	\end{pmatrix}\begin{pmatrix}
	\ell_1 & \ell_2 & L_1  \\
	-m_1 & m_2 & M_1
	\end{pmatrix} \nonumber \\
& \times \, 8 \pi \, \mathcal I^{\ell_1 \ell_2, L_1 M_1, B}_{\gamma\gamma\gamma} \, , 
\ea
where
\ba \label{IB:ttt}
\mathcal I^{\ell_1 \ell_2, L_1 M_1, B}_{\gamma \gamma \gamma} = \int d \ln k_+ \, d \ln k_- \, & {\cal T}_{\ell_2(t)}^{B}(k_+) \, j_{\ell_1}(k_+ r_L)  \, W_t\left(k_-\right) \, \mathcal A_t(k_+) \, \mathcal A_t(k_-) \, f_{L_1, M_1}^{\rm t t t}  \nonumber \\
&\times \left[ (\xi_{LLL} + \xi_{LLR}) - (-1)^{\ell_1 + \ell_2 + L_1} \,(\xi_{RRL} + \xi_{RRR}) \right] \, .
\ea

\subsection{Comments} \label{sec:comments}

Let us make a few comments about these results. It turns out that $\langle \mu_{\ell_1}  X_{\ell_2}\rangle$ angular cross-correlations may get $m$-dependent off-diagonal values ($\ell_1 \neq \ell_2$) as a result of statistical anisotropies (induced by the long-mode $\hat k_l$ angular dependence) introduced in the squeezed primordial bispectra defined in sec. \ref{sec:non-gaus}.

By virtue of the angular momentum algebra of the Wigner symbols, non-vanishing signals are limited to $|\ell_1 - \ell_2| \leq L_1$. For analogous reasons, in cases where the long-wavelength mode is a tensor ($\langle\zeta \zeta \gamma \rangle$ and $\langle \gamma \gamma \gamma \rangle$), a non-zero signature requires $L_1 \geq 2$. This means that by introducing the $\hat k_l$ dependence, we generate at least quadrupolar statistical anisotropies. Moreover, depending by the way in which primordial bispectra transform under parity transformation, only a given $\ell_1$, $\ell_2$ doublet can get a non-zero contribution. By applying the property of the Wigner symbols \eqref{eq:lwigner} we can easily verify the following identities
\begin{align}
C^{\mu X, \,\,\zeta \zeta \zeta}_{\ell_1 \ell_2 m_1 m_2} =& (-1)^{\ell_1 + \ell_2 + L_1} \, C^{\mu X, \,\,\zeta \zeta \zeta}_{\ell_1 \ell_2 m_1 m_2}    \, , \\
C^{\mu X, \,\,\gamma \gamma \zeta}_{\ell_1 \ell_2 m_1 m_2} =& (-1)^{\ell_1 + \ell_2 + L_1} \, C^{\mu X, \,\,\gamma \gamma  \zeta}_{\ell_1 \ell_2 m_1 m_2} \,,
\end{align}
with $X = T, E$. Therefore, by collecting together what we found in this section and eq.~\eqref{eq:Aparityeven}, it is straightforward to realize that for the $\langle \zeta \zeta \zeta \rangle$ bispectrum and bispectra involving tensors with parity-even polarization coefficients we get a non-zero contribution in $\ell_1, \ell_2$ doublets satisfying
\be
\ell_1+\ell_2+L_1 = \begin{cases}
\mbox{even \qquad when  } $X = E, T$\\
\mbox{odd \qquad when  } $X = B$ \, .
\end{cases}
\ee
The same holds for the $\langle \gamma \gamma \zeta \rangle$ bispectrum with maximum violation of parity as described in eq. \eqref{eq:Anoparity}. Moreover, a $\langle \gamma \gamma \zeta \rangle$ bispectrum with parity-odd polarization coefficients would leave no signatures as $\sum_{\lambda \lambda'}  \xi_{\lambda \lambda'} = 0$. On the other hand, by eq. \eqref{eq:Aparityodd} it follows that for $\langle \zeta \zeta \gamma \rangle$ and $\langle \gamma \gamma \gamma \rangle$ bispectra with parity-odd polarization coefficents a non-zero signal is confined to
\be
\ell_1+\ell_2 +L_1= \begin{cases}
\mbox{odd \qquad when  } $X = E, T$\\
\mbox{even \qquad when  } $X = B$ \, .
\end{cases}
\ee
 Finally,  no general conditions (apart for the constraint $|\ell_1 - \ell_2| \leq L_1$) apply to $\langle \zeta \zeta \gamma \rangle$ and $\langle \gamma \gamma \gamma \rangle$ bispectra with maximum parity violation. 

From our explicit calculations we note that the $\hat k_- (\equiv \hat k_s)$ dependence in the angular integrations is always through the spin-0 spherical harmonics $Y_{L_2 M_2}(\hat{k}_-)$ only (see e.g. eq. \eqref{eq:ks_zero}). Therefore, the resultant angular integration over $\hat{k}_-$ is always zero unless $L_2 =0$, i.e. in absence of a $\hat k_- $ angular dependence. It follows that statistical anisotropies induced by the $\hat k_s$ dependence (and labelled by $L_2 \neq 0$) get erased and do not contribute to SD-CMB cross-correlations. The physical interpretation of this comes from the physics of the spectral distortions: when we compute the $\mu$-distortion from the dissipation of acoustic-waves, we need to average the effect of primordial perturbations inside a spherical shell around the last scattering surface with a radius of order the dissipation scale at recombination (see, e.g., \cite{Pajer:2012vz}). As a consequence, any $\hat k_s$ explicit angular dependence is averaged out to zero. 
For the same reason a $\hat k_s$ angular dependence induced by a long tensor mode in a rotationally invariant squeezed bispectrum is erased when averaging over this same spherical shell. Therefore, isotropic squeezed $\langle \gamma \zeta \zeta \rangle$ and $\langle \gamma \gamma \gamma \rangle$ bispectra leave no signatures to SD-CMB cross-correlations. 
This motivates a-posteriori our decision to study the statistically anisotropic case as we are mostly interested on signatures from bispectra involving tensor perturbations.

We end this section by noting that a similar less general discussion was first pointed out in \cite{Shiraishi:2015lma}, where the authors found that diagonal and off-diagonal $\langle \mu T \rangle$ cross-correlations with $|\ell_1 − \ell_2| = 2$ arise in scalar bispectra with a quadrupolar asymmetry (corresponding to our $\langle \zeta \zeta \zeta \rangle$, $L_1 = 2$, $L_2 = 0$ case).

In the next section, we aim to quantify the detectability prospects of the signatures studied in this section.

\section{Forecasts} \label{sec:forecast}

In this section, we make Fisher forecasts on the detectability of statistical anisotropies in primordial NGs with the cross-correlations we have computed in sec. \ref{sec:comp_non_gaus}. We will look into both parity preserving and parity violating patterns. As shown above, statistical anisotropies in squeezed bispectra could lead to off-diagonal elements in the SD-CMB cross-correlations $C^{\mu T}_{\ell_1 \ell_2}$, $C^{\mu E}_{\ell_1 \ell_2}$ and $C^{\mu B}_{\ell_1 \ell_2}$. 

Such statistical anisotropies are most effectively analyzed with the so-called BipoSH formalism~\cite{Hajian:2003qq,Souradeep:2003qr,Hajian:2005jh}. Here we give a brief description of this formalism, referring to the original literature for more details. We begin by considering a generic cross-correlation in  real space between two observables $\mathcal O^1$ and $\mathcal O^2$
\be
\langle \mathcal O^1(\hat x_1) \mathcal O^2(\hat x_2) \rangle \, ,
\ee
where $\hat x_1$ and $\hat x_2$ correspond to two different directions in the sky. We can expand this quantity as
\be \label{eq:obs_anis}
\langle \mathcal O^1(\hat x_1) \mathcal O^2(\hat x_2) \rangle = \sum_{\ell_1 \ell_2, L M} A_{\ell_1 \ell_2}^{L M, \mathcal O^1 \mathcal O^2} \left\{ Y_{\ell_1}(\hat x_1) \otimes Y_{\ell_2}(\hat x_2) \right\}_{L M}\, ,
\ee
where we have introduced the bipolar spherical harmonics
\be
\left\{ Y_{\ell_1}(\hat x_1) \otimes Y_{\ell_2}(\hat x_2) \right\}_{L M} = \sum_{m_1  m_2} \mathcal C^{L M}_{\ell_1 m_1 \ell_2 m_2} Y_{\ell_1 m_1}(\hat x_1) Y_{\ell_2 m_2}(\hat x_2) \, .
\ee
The quantities $\mathcal C^{L M}_{\ell_1 m_1 \ell_2 m_2}$ are the so-called Clebsch-Gordan coefficients (see app. \ref{app:Wigner}). By inverting eq. \eqref{eq:obs_anis} and doing the angular integrations we get the BipoSH coefficients
\be \label{def:bip}
A_{\ell_1 \ell_2}^{L M, \mathcal O^1 \mathcal O^2} = \sum_{m_1  m_2}  (-1)^{m_2} \, \mathcal C^{L M}_{\ell_1 m_1 \ell_2 - m_2} \, \langle \mathcal O^1_{\ell_1 m_1} \mathcal O^2_{\ell_2 m_2} \rangle \, .
\ee
When statistical isotropy holds, the BipoSH coefficients vanish for $L > 0$ and for $L = 0$ we recover the usual diagonal angular correlations
\be
A_{\ell_1 \ell_2}^{0 0, \mathcal O^1 \mathcal O^2} = \delta_{\ell_1 \ell_2} (-1)^{\ell_1} (2 \ell_1 + 1)^{1/2} \, \langle \mathcal O^1_{\ell_1} \mathcal O^2_{\ell_2} \rangle \, .
\ee
On the other hand, when statistical isotropy is broken, we can use \eqref{def:bip} for $L>0$ to characterize the anisotopies. In particular, we can build the following unbiased estimator for the BipoSH coefficients
\be \label{def:bip_est}
\hat A_{\ell_1 \ell_2}^{L M, \mathcal O^1 \mathcal O^2} = \sum_{m_1  m_2} (-1)^{m_2} \, \mathcal C^{L M}_{\ell_1 m_1 \ell_2 - m_2} \, \mathcal O^1_{\ell_1 m_1} \mathcal O^2_{\ell_2 m_2} \, .
\ee
Assuming it depends on the parameters $\theta_i$ and $\theta_j$ of an underlying theory, we can define the resultant Fisher-matrix as
\be \label{eq:Fisher_anis}
F_{\theta_i, \theta_j} = \frac{\partial A_{\ell_1 \ell_2}^{L M, \mathcal O^1 \mathcal O^2}}{\partial \theta_i} \frac{\partial A_{\ell'_1 \ell'_2}^{* L' M', \mathcal O^1 \mathcal O^2}}{\partial \theta_j} \left(C^{-1}_{A A^*}\right)_{ij} \, ,
\ee
where the covariance matrix reads
\be \label{eq:cov_anis}
C_{A A^*} = \langle \hat A_{\ell_1 \ell_2}^{L M, \mathcal O^1 \mathcal O^2} \hat A_{\ell'_1 \ell'_2}^{* L' M', \mathcal O^1 \mathcal O^2}\rangle \, .
\ee
In the following we will use these BipoSH coefficients and eq. \eqref{eq:Fisher_anis} to make Fisher forecasts on the detectability of statistical anisotropies in primordial (scalar and tensor) NGs in SD-CMB cross-correlations.

\subsection{3-scalars bispectrum}

The SD-CMB cross-correlators sensitive to this primordial bispectrum are $C^{\mu T}_{\ell_1 \ell_2 m_1 m_2}$ and $C^{\mu E}_{\ell_1 \ell_2 m_1 m_2}$ angular cross-spectra. By substituting \eqref{eq:mT_sss_final} and \eqref{eq:mE_sss_final} into \eqref{def:bip} and employing the properties of the Wigner symbols (refer to eq. \eqref{eq:sum_m_wigner} of app. \ref{app:Wigner}) we get the following BipoSH coefficients
\ba \label{eq:A_sssT}
A_{\ell_1 \ell_2}^{L M, \,\mu T} = &  \, \delta_{M M_1} \,\, i^{\ell_1-\ell_2} \,  \sqrt{(2 \ell_1 + 1) (2 \ell_2 + 1)} \, \begin{pmatrix}
	\ell_1 & \ell_2 & L \\
	0 & 0 & 0
	\end{pmatrix} \nonumber \\
& \times \, 4 \pi \, \mathcal I^{\ell_1 \ell_2, L M, T}_{\zeta \zeta \zeta } \, ,
\ea
and
\ba \label{eq:A_sssE}
A_{\ell_1 \ell_2}^{L M, \,\mu E} = &\, \,\, i^{\ell_1-\ell_2} \, \sqrt{(2 \ell_1 + 1) (2 \ell_2 + 1)} \, \begin{pmatrix}
	\ell_1 & \ell_2 & L \\
	0 & 0 & 0
	\end{pmatrix} \nonumber \\
& \times \, 4 \pi \, \mathcal I^{\ell_1 \ell_2, L M, E}_{\zeta \zeta \zeta }  \, ,
\ea
where $\mathcal I^{\ell_1 \ell_2,L M, T}_{\zeta \zeta \zeta}$ and $\mathcal I^{\ell_1 \ell_2,L M, E}_{\zeta \zeta \zeta}$ are as in eqs. \eqref{IT:sss} and \eqref{IE:sss}.

The variance of the quantities just introduced reads
\ba
\sigma^2(A_{\ell_1 \ell_2}^{LM, \,\mu X}) = \sum_{m_1 m_1'}\,\sum_{m_2 m_2'}  & \left[C^{\mu \mu}_{\ell_1  \ell_1 m_1 m_1'} \,C^{XX}_{\ell_2 \ell_2 m_2 m_2'}  \, + \, C^{\mu X}_{\ell_1  \ell_2 m_1 m_2'} \,C^{X \mu}_{\ell_2 \ell_1 m_2 m_1'} \right] \\
& \times (-1)^{m_2 + m_2'}\mathcal C^{L M}_{\ell_1 m_1 \ell_2 -m_2} \, \mathcal C^{L M}_{\ell_1 m_1' \ell_2 - m_2'} \simeq  C^{\mu \mu}_{\ell_1} \, C^{XX}_{\ell_2} \, ,
\ea
where $X= T, E$ and the last approximation holds in the regime of small primordial NGs, where we should expect
\be
\left(C^{\mu X}_{\ell m}\right)^2 \ll C^{X X}_{\ell m} C^{\mu \mu}_{\ell m} \, . 
\ee
Here the $C^{X X}_{\ell m}$'s are the CMB total $X$-mode power spectra, which we will assume to be cosmic-variance limited on large scales. Moreover, given the current and planned experiments aiming to measure the $\mu$-spectral distortions of the CMB, we expect that the experimental noise in the $\mu$ modes angular power spectrum dominates over the signal, i.e. $C^{\mu \mu}_{\ell m, \rm N} \gg C^{\mu \mu}_{\ell m, \rm sign}$ \footnote{See \cite{Pajer:2012vz} for more details in this regards. Needless to say, assuming we can build a (very futuristic) experiment where we can make a cosmic-variance limited measurement of $\mu$-distortions, a lot of further improvement in the detection of squeezed bispectra should be expected, in line with ref. \cite{Kalaja:2020mkq}. However, here we are not focusing in this scenario and consider noise of planned CMB experiments.}.

For a PIXIE-like experiment the expected level of noise is given by \cite{Kogut:2011xw,Kogut:2019vqh}
\be
C^{\mu \mu, \rm PIXIE}_{\ell m, \rm N} =  \mu_N^2 \times e^{(\ell/84)^2} \, , %C^{\mu \mu, \rm Super-PIXIE}_{\ell m, \rm N} =4 \pi \times (7.7 \times 10^{-9})^2 \times e^{(\ell/84)^2} \, .
\ee
where $\mu_N = 4.96 \times 10^{-8}$. %Here we did not account for the contribution of galactic foregrounds. As also noticed in \cite{Abitbol:2017vwa,Remazeilles:2018kqd}, these should be taken into account when estimating the level of degradation in the $S/N$ ratio for a real world experiment.
Under these assumptions, we get
\ba
\sigma^2(A_{\ell_1 \ell_2}^{LM, \,\mu X})  \simeq  C^{\mu \mu}_{\ell_1, \rm N} \, C^{XX}_{\ell_2} \, .
\ea
Therefore, we get the following Fisher matrix for the parameter $f^{sss}_{L} = f^{sss}_{L, M} $ from $A_{\ell_1 \ell_2}^{L M, \,\mu X}$, \footnote{Here and afterwards we drop the $M$ dependences on the coefficients $f^{xxx}_{LM}$'s as the forecasts do not depend by $M$.}
\be
F_L = \sum_{\ell_1, \ell_2=2}^{\ell_{\rm max}}  \frac{\tilde A_{\ell_1 \ell_2}^{L M, \,\mu X} \, \tilde A_{\ell_1 \ell_2}^{* L M, \,\mu X}}{C^{\mu \mu}_{\ell_1, \rm N} \, C^{XX}_{\ell_2} } \, ,
\ee
where $\tilde A_{\ell_1 \ell_2}^{L M, \,\mu X} =  \partial A_{\ell_1 \ell_2}^{L M, \,\mu X}/\partial f^{sss}_{L}$. Assuming that our observables are Gaussian distributed the expected 1-sigma error on $f_{L}^{sss}$ is given by
\be
\Delta f_{L}^{sss} = F_L^{-1/2} \, .
\ee
%However, We stress that this represents only the lower bound on the error, and it corresponds to the exact error only when our observables are Gaussian distributed. In our case the angular cross-correlators follow a $\chi^2$ distribution with $2 \ell + 1$ degrees of freedom, which approaches a Gaussian distribution only in the large-$\ell$ limit (see, e.g., \cite{Joshi:2011vc}).
If we want to combine the $T$ and $E$ modes in our estimate, we need to write down the following joint-Fisher matrix
\be \label{Fisher:TE}
F_L = \sum_{\ell_1, \ell_2=2}^{\ell_{\rm max}} \,  \mathbf{\tilde A}^{L M}_{\ell_1 \ell_2} \cdot C_{A A*,\, \ell_1 \ell_2}^{-1} \cdot \mathbf{\tilde A}^{* L M, T}_{\ell_1 \ell_2}  \, ,
\ee
where 
\be
\mathbf{\tilde A}^{L M}_{\ell_1 \ell_2} = \begin{pmatrix}
	 \tilde A^{L M, \mu T}_{\ell_1 \ell_2}\\
                                          \\
	 \tilde A^{L M, \mu E}_{\ell_1 \ell_2} 
	\end{pmatrix}  \, , \qquad C_{A A*,\, \ell_1 \ell_2} \simeq \begin{pmatrix}
	 C^{\mu \mu}_{\ell_1, \rm N} C^{TT}_{\ell_2} & \,\,C^{\mu \mu}_{\ell_1, \rm N} C^{TE}_{\ell_2} \\
                                          \\
	 C^{\mu \mu}_{\ell_1, \rm N} C^{TE}_{\ell_2} & \,\,C^{\mu \mu}_{\ell_1, \rm N} C^{EE}_{\ell_2} 
	\end{pmatrix} \, .
\ee
Therefore
\be 
C^{-1}_{A A*,\, \ell_1 \ell_2} \simeq \begin{pmatrix}
	 \frac{C^{EE}_{\ell_2}}{C^{\mu \mu}_{\ell_1, \rm N} \left(  C^{TT}_{\ell_2} C^{EE}_{\ell_2} - (C^{TE}_{\ell_2})^2 \right)}  & \,\,  - \frac{C^{TE}_{\ell_2}}{C^{\mu \mu}_{\ell_1, \rm N} \left(  C^{TT}_{\ell_2} C^{EE}_{\ell_2} - (C^{TE}_{\ell_2})^2 \right)} \\
                                          \\
	 - \frac{C^{TE}_{\ell_2}}{C^{\mu \mu}_{\ell_1, \rm N} \left(  C^{TT}_{\ell_2} C^{EE}_{\ell_2} - (C^{TE}_{\ell_2})^2 \right)} & \,\,\frac{C^{TT}_{\ell_2}}{C^{\mu \mu}_{\ell_1, \rm N} \left(  C^{TT}_{\ell_2} C^{EE}_{\ell_2} - (C^{TE}_{\ell_2})^2 \right)} 
	\end{pmatrix} \, .
\ee
%\be
%F_L = \sum_{\ell_1, \ell_2=2}^{\ell_{\rm max}} \, \frac{C^{TT}_{\ell m} \left|\tilde A^{L M, \mu E}_{\ell_1 \ell_2}\right|^2 + C^{EE}_{\ell_2 m} \left|\tilde A^{L M,\mu T}_{\ell_1 \ell_2}\right|^2 - 2  C^{TE}_{\ell_2} \left(\tilde A^{L M, \mu T}_{\ell_1 \ell_2} \tilde A^{* L M, \mu E}_{\ell_1 \ell_2}\right)}{C^{\mu \mu}_{\ell_1, \rm N} \left[C^{TT}_{\ell_2} C^{EE}_{\ell_2} - (C^{TE}_{\ell_2})^2 \right]} \, .
%\ee
It is worth to stress that our expressions for the Fisher matrix are valid when considering a full-sky experiment. In a real world experiment, the F-matrix is damped by a factor $f_{\rm sky}$, where $f_{\rm sky}$ is the portion of the sky covered by a given CMB survey.
\begin{figure}[H] 
\centering
   \includegraphics[width=.48\textwidth]{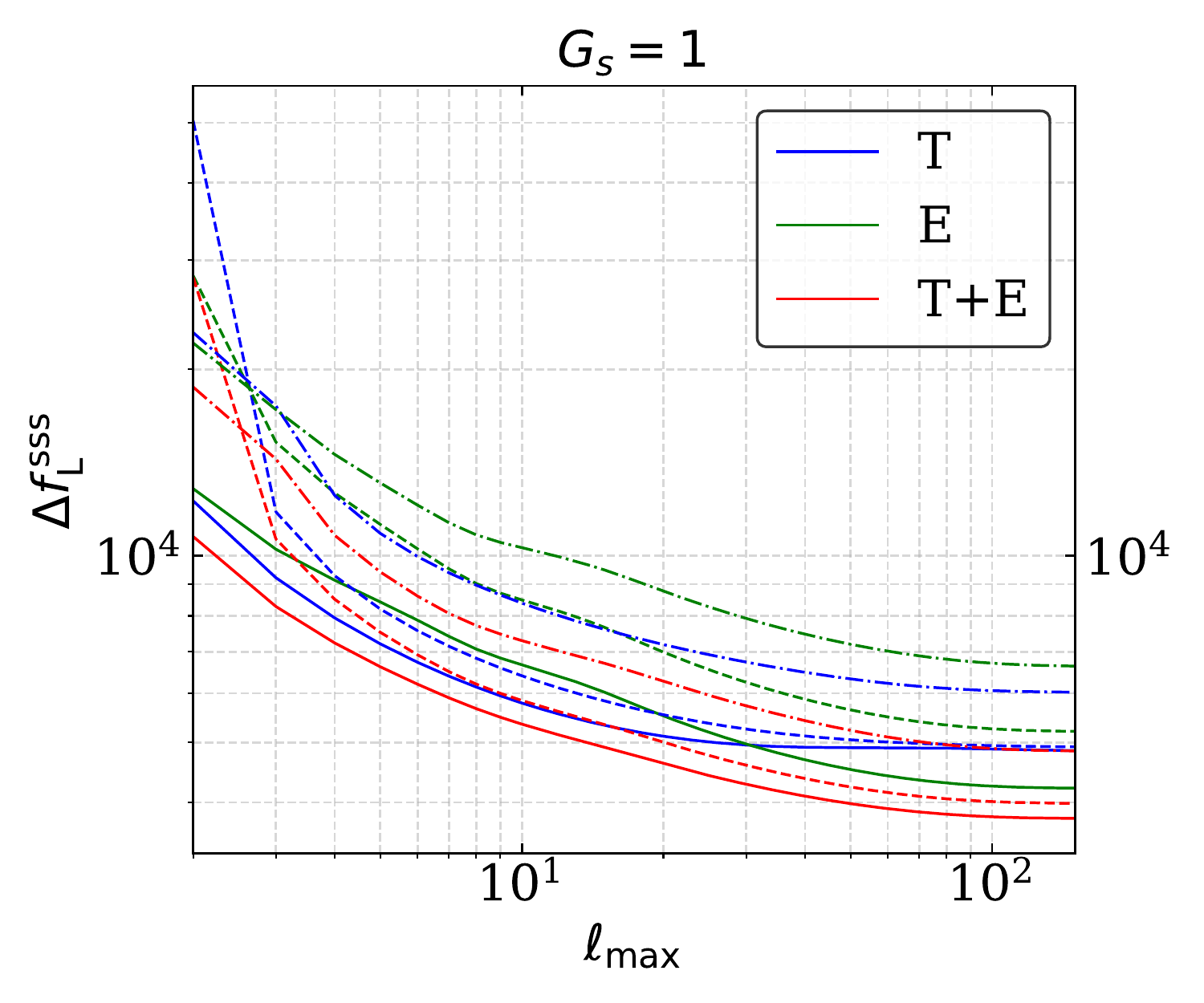} \,\, \includegraphics[width=.48\textwidth]{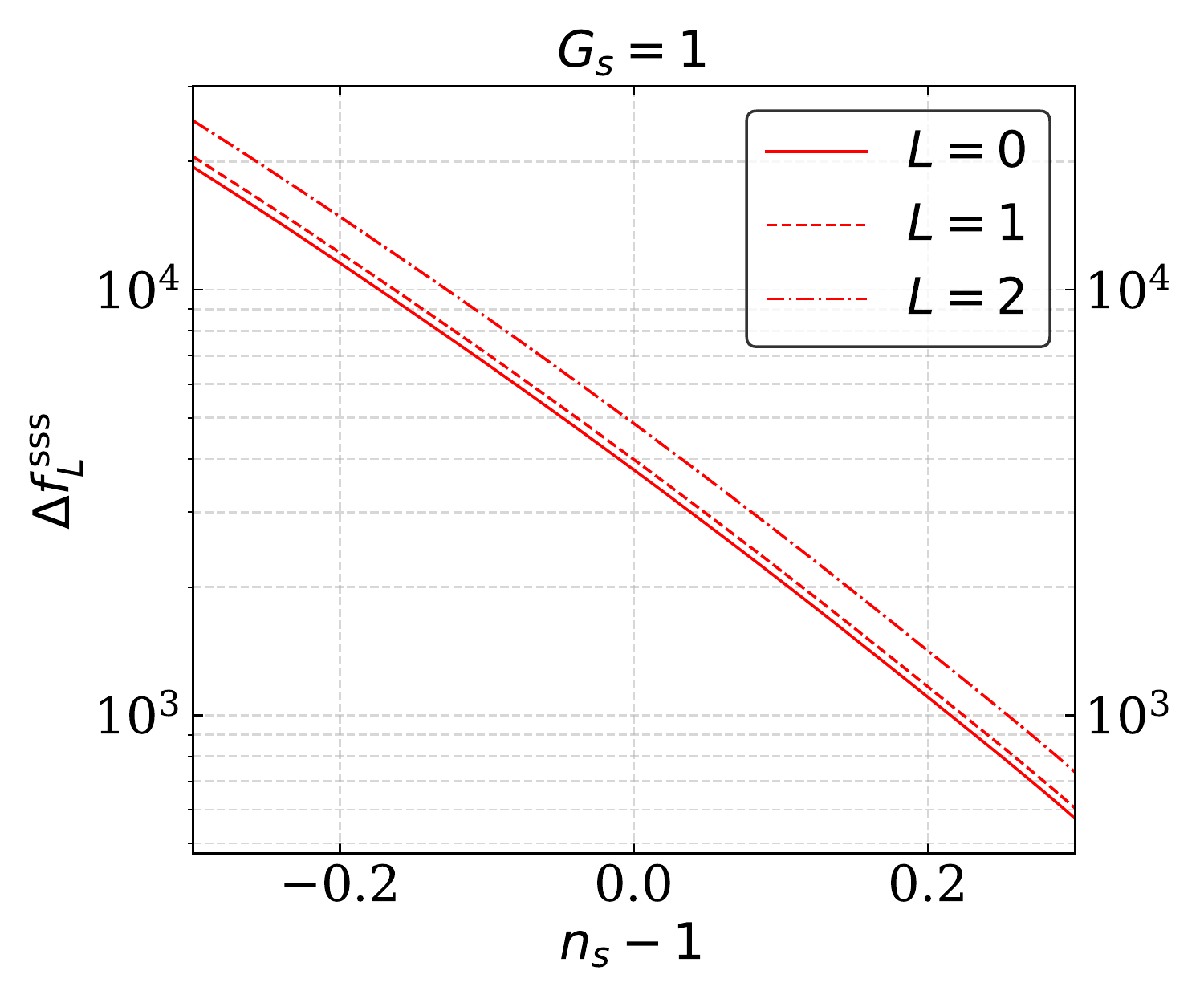}
  \caption{Plot of the expected 1-sigma error on $f^{\rm sss}_L$ from $\mu T$ and $\mu E$ for a PIXIE-like level of noise. Plain lines corresponds to $L = 0$, dashed lines to $L = 1$, dot-dashed lines to $L = 2$. Left Panel: $\Delta f_L^{\rm sss}$ vs $\ell_{\rm max}$ for scale invariant scalar power spectrum. Right panel: $\Delta f_{L}^{\rm sss}$ vs $n_s - 1$. 
  \label{fig:Fishersss}}
\end{figure}
In fig. \ref{fig:Fishersss} we plot the expected 1-sigma error on  $f_{L}^{sss}$ for different kind of statistical anisotropies labeled by $L = 0, 1, 2$ obtained combining the $C^{\mu T}_{\ell m}$ and $C^{\mu E}_{\ell m}$ angular cross-spectra. 
The plots are made taking $G_s = 1$ and varying the scalar-tilt at the SD scales. From our numerical results we obtain the following scaling formula \footnote{Here and afterwards the scalar and tensor tilts dependence is the result of fits in the $n_s-1$ ($n_t$) region space $[-0.3,0.3]$ (the relative errors are within $1\%$). We verified that, by modifying this region, the dependence over $n_s$ is not significantly altered (the relative errors stay within $10\%$ for $|n_s-1| \leq 1$). In contrary, the dependence over $n_t$ is much more sensitive to the tensor-tilt region considered, so the corresponding fits should be considered as very rough estimates. This sensitivity to $n_t$ is due to the soft decay at $k>10^4 \, \mbox{Mpc}^{-1}$ of the tensor SD-transfer function, allowing for increasing scales to contribute in the integrals as \eqref{IT:tts} with increasing values of $n_t$.}
\ba \label{eq:Delta_sss_0}
\Delta f_{L}^{sss}|_{T+E} = a  \, 10^{b(n_s-1) + c(n_s-1)^2} \, .
\ea
Analyzing the expression for the Fisher matrix \eqref{Fisher:TE}, this formula can be generalized to $G_s \neq 1$ and with generic level of noise $\mu_N$ as
\ba \label{eq:Delta_sss}
\Delta f_{L}^{sss}|_{T+E} = \frac{a}{G_s} \, \left(\frac{\mu_{\rm N}}{4.96 \times 10^{-8}}\right) \, 10^{b(n_s-1) + c(n_s-1)^2} \, .
\ea
In tab. \ref{tab:fit_power_sss} we summarize the values of the fit parameters $a$, $b$, $c$ for the various $L$-poles. 
\begin{table}[H]
\begin{center}
\begin{tabular}{ |c|c|c|c| } 
\hline
$L$ & $a$ & $b$ & $c$ \\ 
\hline 
0 & $3.7 \times 10^3$ & $-2.55$ & $-0.58$ \\ 
\hline
1 & $3.9 \times 10^3$ & $-2.55$ & $-0.58$ \\ 
\hline
2 &  $4.8 \times 10^3$ &  $-2.55$ & $-0.58$ \\
\hline
\end{tabular}
\end{center}
\caption{Values of the parameters in eq. \eqref{eq:Delta_sss}.} \label{tab:fit_power_sss}
\end{table} 
Our results suggest that the detectability prospects decrease by increasing the $L$-pole labelling a given statistical anisotropy. However, by admitting statistical anisotropies with $L \leq 2$, the detectability prospects remain commensurate.

\subsection{2-tensors 1-scalar bispectrum}

\begin{figure}[H]  
\centering
   \includegraphics[width=.48\textwidth]{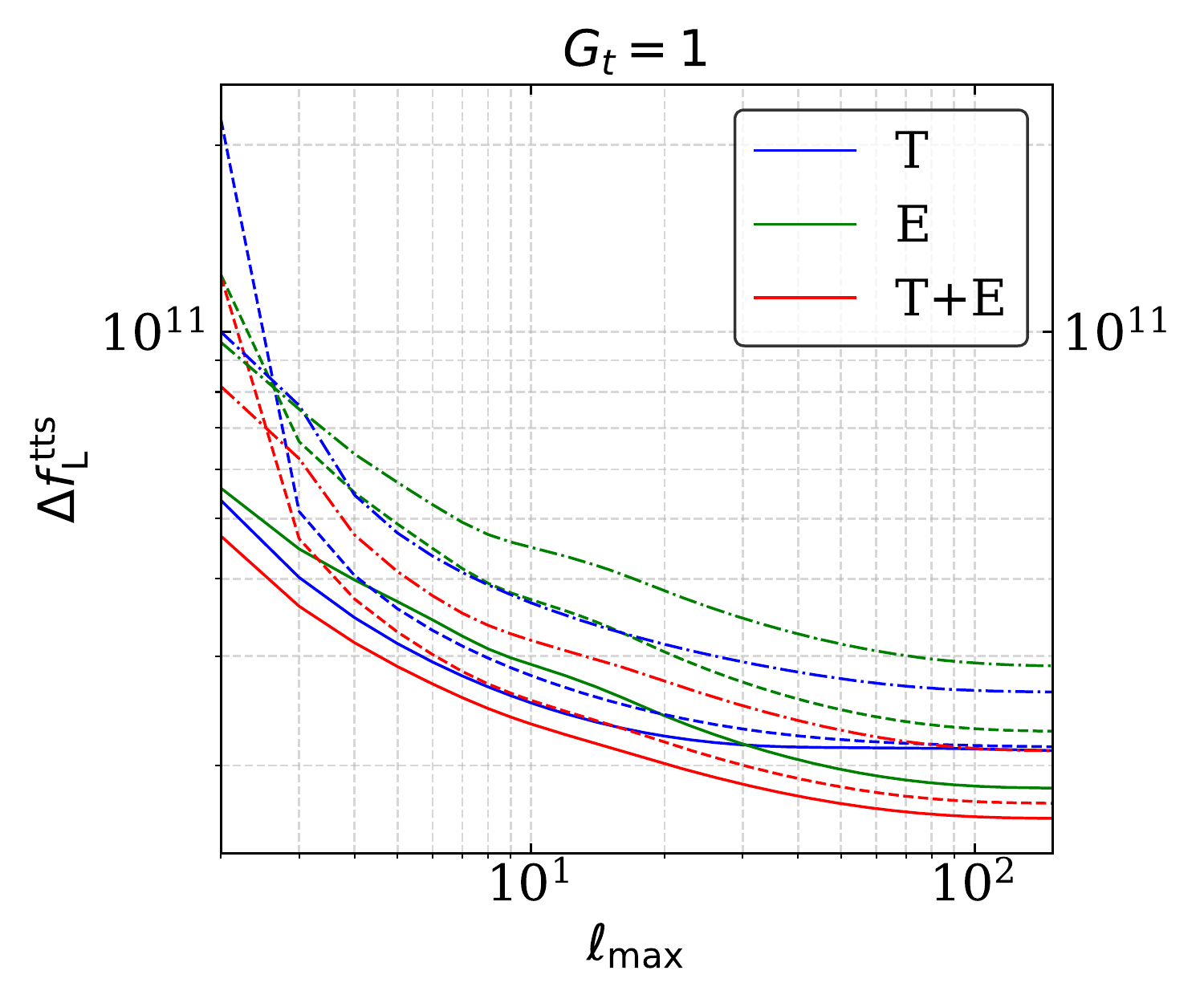} \,\, \includegraphics[width=.48\textwidth]{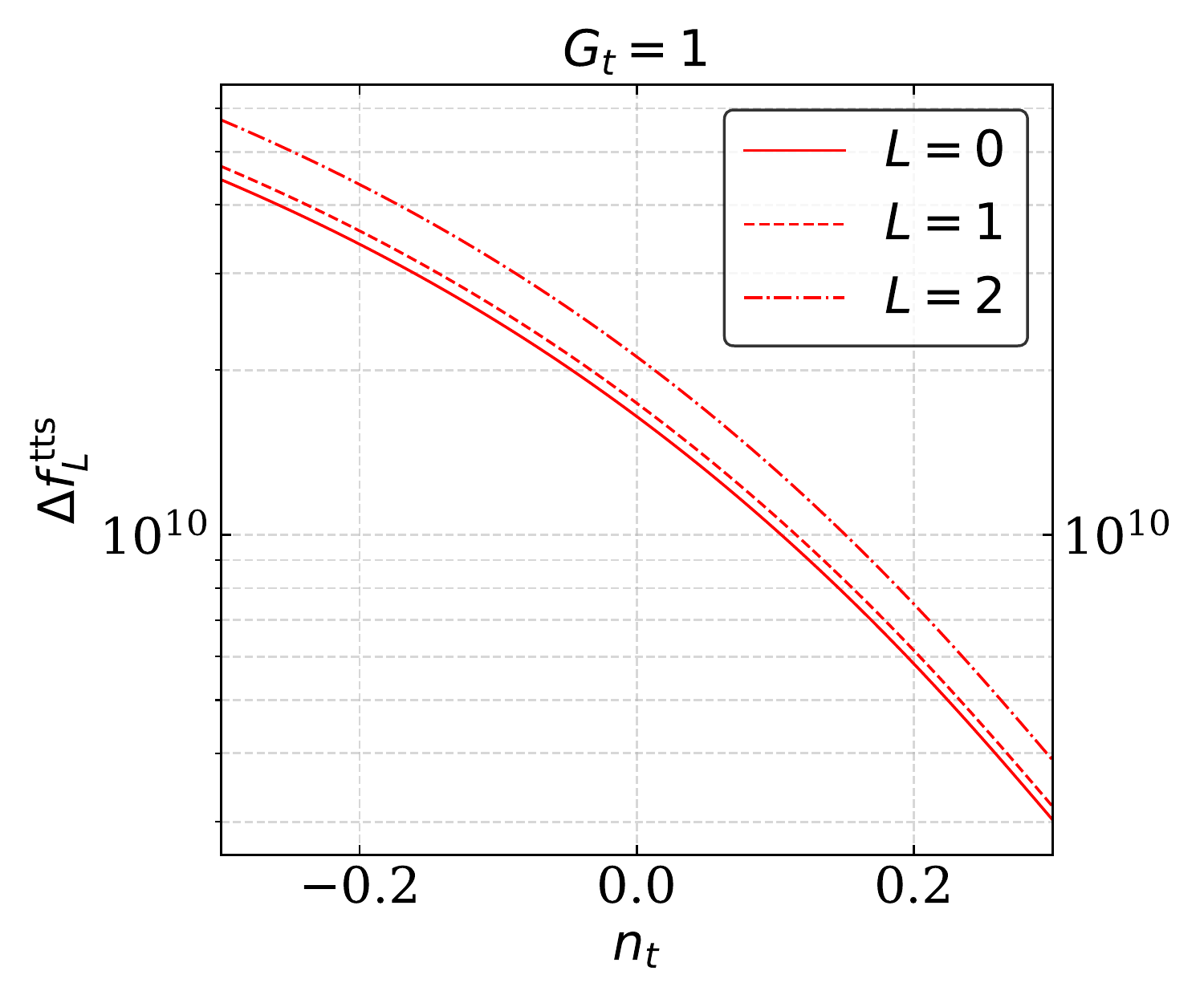}
  \caption{Plot of the expected 1-sigma error on $f^{\rm tts}_L$ from $\mu T$ and $\mu E$ for PIXIE-like noise. The tensor-to-scalar ratio at CMB scales is taken $r_{\rm CMB} = 0.01$. Plain lines corresponds to $L = 0$, dashed lines to $L = 1$, dot-dashed lines to $L = 2$. Left Panel: $\Delta f_L^{\rm tts}$ vs $\ell_{\rm max}$ for scale invariant tensor power spectrum. Right panel: $\Delta f_{L}^{\rm tts}$ vs $n_t$. 
  \label{fig:Fishertts}}
\end{figure}
As in the previous subsection, SD-CMB cross-correlators sensitive to the 2-tensors 1-scalar primordial bispectrum are $C^{\mu T}_{\ell_1 \ell_2 m_1 m_2}$ and $C^{\mu E}_{\ell_1 \ell_2 m_1 m_2}$ angular cross-spectra. By substituting \eqref{eq:mT_tts_final} and \eqref{eq:mE_tts_final} into \eqref{def:bip} we get
\ba \label{eq:A_ttsT}
A_{\ell_1 \ell_2}^{L M, \,\mu T} = &  \,  \,\, i^{\ell_1-\ell_2}  \,  \sqrt{(2 \ell_1 + 1) (2 \ell_2 + 1)} \, \begin{pmatrix}
	\ell_1 & \ell_2 & L \\
	0 & 0 & 0
	\end{pmatrix} \nonumber \\
& \times \, 8 \pi \, \left(\sum_{\lambda \lambda'}  \xi_{\lambda \lambda'}\right) \, \mathcal I^{\ell_1 \ell_2, L M, T}_{\gamma \gamma \zeta } \, ,
\ea
and
\ba \label{eq:A_ttsE}
A_{\ell_1 \ell_2}^{L M, \,\mu E} = &\,  \,\, i^{\ell_1-\ell_2}  \, \sqrt{(2 \ell_1 + 1) (2 \ell_2 + 1)} \, \begin{pmatrix}
	\ell_1 & \ell_2 & L \\
	0 & 0 & 0
	\end{pmatrix} \nonumber \\
& \times \, 8 \pi \, \left(\sum_{\lambda \lambda'} \xi_{\lambda \lambda'}\right) \, \mathcal I^{\ell_1 \ell_2, L M, E}_{\gamma \gamma \zeta } \, ,
\ea
where $\mathcal I^{\ell_1 \ell_2, L M, T}_{\gamma \gamma \zeta}$ and $\mathcal I^{\ell_1 \ell_2, L M, E}_{\gamma \gamma \zeta}$ are as in eqs. \eqref{IT:tts} and \eqref{IE:tts}. Here we adopt the convention $|\xi_{RR}| = |\xi_{LL}|=1$, and we assume $\xi_{LR} = \xi_{RL} = 0$ \footnote{We are not considering the contribution of squeezed $\langle \gamma \gamma \zeta \rangle$ bispectra that involve mixed chiralities. These bispectra are typically very sensitive to the details and the symmetry breaking patterns of the underlying model (see e.g. \cite{Dimastrogiovanni:2018gkl,Bartolo:2020gsh}). However, we would expect $\langle \gamma_R \gamma_L \zeta \rangle$ and $\langle \gamma_L \gamma_R \zeta \rangle$ to give a signature at most of the same order of magnitude than $\langle \gamma_R \gamma_R \zeta \rangle$ and $\langle \gamma_L \gamma_L \zeta \rangle$, leading to an improvement of only a factor 2 in the 1-sigma error on $f_L^{tts}$.}. In fig. \ref{fig:Fishertts} we plot the expected 1-sigma error on $f^{tts}_{L}$ for $L =0, 1, 2$. Using the plots and the expression of the Fisher matrix we can get the following scaling formula of the expected 1-sigma error
\ba \label{eq:Delta_tts}
\Delta f_{L}^{tts}|_{T+E} = \frac{a}{G_t} \, \left(\frac{\mu_{\rm N}}{4.96 \times 10^{-8}}\right) \, \left(\frac{0.01}{r_{\rm CMB}}\right) 10^{b(n_t) + c(n_t)^2} \, ,
\ea
where $r_{\rm CMB}$ denotes the tensor-to-scalar ratio at CMB scales, and the values of the parameters are given in tab. \ref{tab:fit_power_tts}.
\begin{table}[h!]
\begin{center}
\begin{tabular}{ |c|c|c|c| } 
\hline
$L$ & $a$ & $b$ & $c$ \\ 
\hline
0 & $1.6 \times 10^{10}$  & $-1.91$ & $-1.52$ \\
\hline
1 & $1.7 \times 10^{10}$ & $-1.91$ & $-1.52$ \\ 
\hline
2 &  $2.1 \times 10^{10}$ &  $-1.91$ & $-1.52$ \\
\hline
\end{tabular}
\end{center}
\caption{Values of the parameters in eq. \eqref{eq:Delta_tts}.} \label{tab:fit_power_tts}
\end{table}
Again, we notice a degradation in the detection prospects with increasing levels of statistical anisotropies, while the dependence on the other relevant parameters is not altered by the kind of statistical anisotropy.

\subsection{2-scalars 1-tensor bispectrum}

Intuitively, the SD-CMB cross-correlation most sensitive to the 2-scalars 1-tensor primordial bispectrum is the $C^{\mu B}_{\ell_1 \ell_2 m_1 m_2}$ cross-spectrum. In fact, cross-correlations of $\mu$ modes with $T$ and $E$ modes generated by tensor perturbations are expected to be limited by the scalar induced cosmic variance-limited $T$- and $E$-mode power spectra. On large scales, the CMB tensor transfer functions ${\cal T}_{\ell(t)}^{X}$ are comparable in size to the scalar transfer functions ${\cal T}_{\ell(s)}^{X}$. Therefore, the $\mu T (\mu E)$ and $\mu B$ Fisher matrices per unit-$\ell$ scale as
\be
\frac{F_L^{\mu T}}{F_L^{\mu B}}\Big|_{\ell} \sim \frac{C_\ell^{BB}}{C_\ell^{TT}} \sim r_{\rm CMB} \, .
\ee
As $r_{\rm CMB} < 0.056$, we have an increase of at least 1 order of magnitude in the 1-sigma error on $f^{sst}_{L}$ by using $\mu T$ and $\mu E$ rather than  $\mu B$. 

By substituting eq. \eqref{eq:mB_sst_final} into eq. \eqref{def:bip} and employing the properties of the Wigner symbols we get
\ba
A_{\ell_1 \ell_2}^{L M, \,\mu B} = & \,\, i^{\ell_1-\ell_2} \,\sqrt{(2 \ell_1 + 1) (2 \ell_2 + 1)} \, \begin{pmatrix}
	\ell_1 & \ell_2 & L \\
	0 & 2 & -2
	\end{pmatrix} \nonumber \\
& \times \, 4 \pi \, \mathcal I^{\ell_1 \ell_2, L M, B}_{\zeta \zeta \gamma} \,  ,
\ea
where $\mathcal I^{\ell_1 \ell_2, L M, B}_{\zeta \zeta \gamma}$ is as in eq. \eqref{IB:sst}. Here we adopt the convention $|\xi_R|=|\xi_L| = 1$.  

\begin{figure}[H]  
\centering
   \includegraphics[width=.48\textwidth]{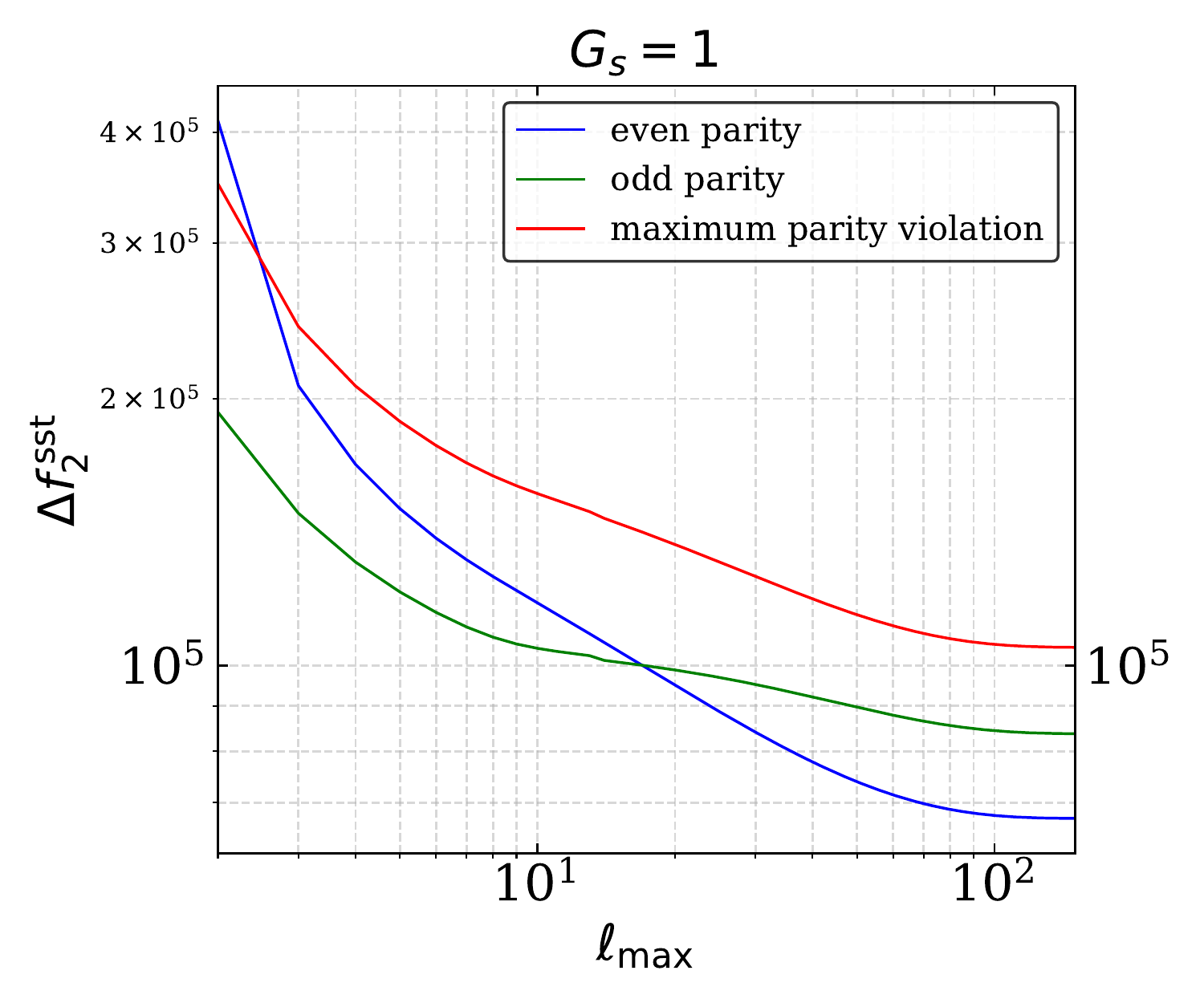} \,\, \includegraphics[width=.48\textwidth]{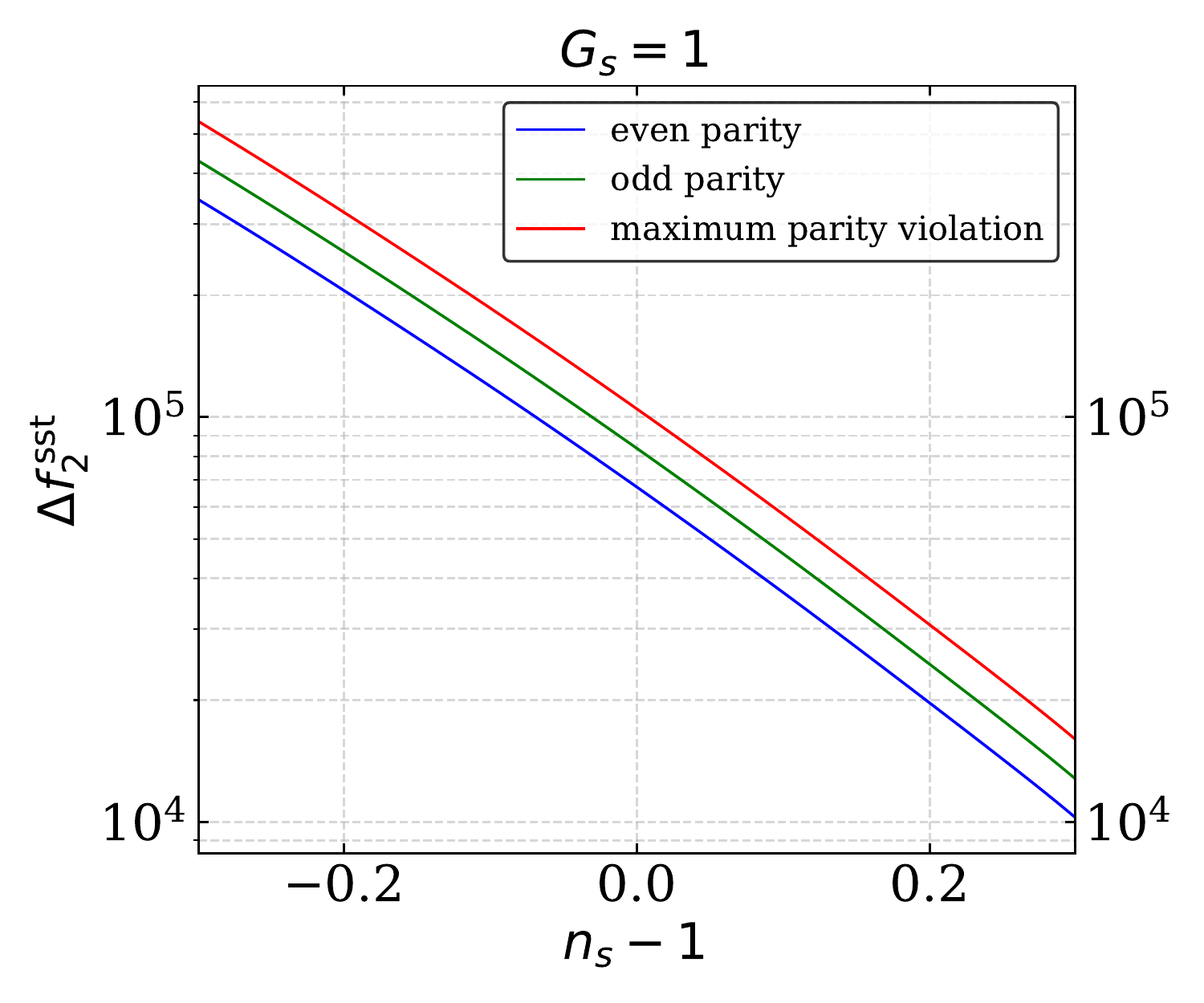}
  \caption{Plot of the expected 1-sigma error on $f^{\rm sst}_2$ from $\mu B$ for PIXIE-like level of noise and cosmic-variance limited fully delensed $B$ modes. The tensor-to-scalar ratio at CMB scales is taken $r_{\rm CMB} = 0.01$. Left Panel: $\Delta f_2^{\rm sst}$ vs $\ell_{\rm max}$ for scale invariant scalar power spectrum. Right panel: $\Delta f_{2}^{\rm sst}$ vs $n_s-1$.
  \label{fig:Fishersst}}
\end{figure}

The computation of the Fisher matrix for $f^{tss}_{L}$ resembles the computations above and we get
\be
F_L = \sum_{\ell_1, \ell_2=2}^{\ell_{\rm max}}  \frac{\tilde A_{\ell_1 \ell_2}^{L M, \,\mu B} \, \tilde A_{\ell_1 \ell_2}^{* L M, \,\mu B}}{C^{\mu \mu}_{\ell_1, \rm N} \, C^{BB}_{\ell_2} } \, ,
\ee
where $\tilde A_{\ell_1 \ell_2}^{L M, \,\mu B} =  \partial A_{\ell_1 \ell_2}^{L M, \,\mu B}/\partial f^{sst}_{L}$.  In fig. \ref{fig:Fishersst} we plot the expected 1-sigma error on $f^{sst}_{L}$, with $L = 2$, for the different ways in which the $\langle \zeta \zeta \gamma \rangle$ bispectrum transforms under parity transformation. 
From the results of the forecasts and the expression of the Fisher matrix, we obtain the following scaling formula
\ba \label{eq:Delta_sst}
\Delta f_{2}^{sst}|_{B} = \frac{a}{G_s} \, \left(\frac{\mu_{\rm N}}{4.96  \times 10^{-8}}\right) \, \left(\frac{0.01}{r_{\rm CMB}}\right)^{1/2} \, 10^{b(n_s-1) + c(n_s-1)^2} \, .
\ea
In tab. \ref{tab:fit_power_sst} we summarize the values of the fit parameters $a$, $b$, $c$. 
\begin{table}[h!]
\begin{center}
\begin{tabular}{ |c|c|c|c| } 
\hline 
Transformation under parity  & $a$ & $b$ & $c$ \\ 
\hline
even parity & $6.8 \times 10^4$ & $- 2.55$ & $-0.58$ \\ 
\hline
odd parity & $8.5 \times 10^4$ & $- 2.55$ & $-0.58$\\ 
\hline
maximum parity violation &  $1.0 \times 10^5$ & $- 2.55$ & $-0.58$ \\
\hline
\end{tabular}
\end{center}
\caption{Values of the parameters in eq. \eqref{eq:Delta_sst}.} \label{tab:fit_power_sst}
\end{table}
Here we have only explored the case of quadrupolar statistical anisotropies ($L = 2$). Higher levels of statistical anisotropies can be studied as well, but we leave such a study for more model dependent settings. We found that detectability prospects slighly degrade when we consider parity violation signatures, even if the final results still remain commensurate. 

\subsection{3-tensors bispectrum}

\begin{figure}[H]  
\centering
   \includegraphics[width=.48\textwidth]{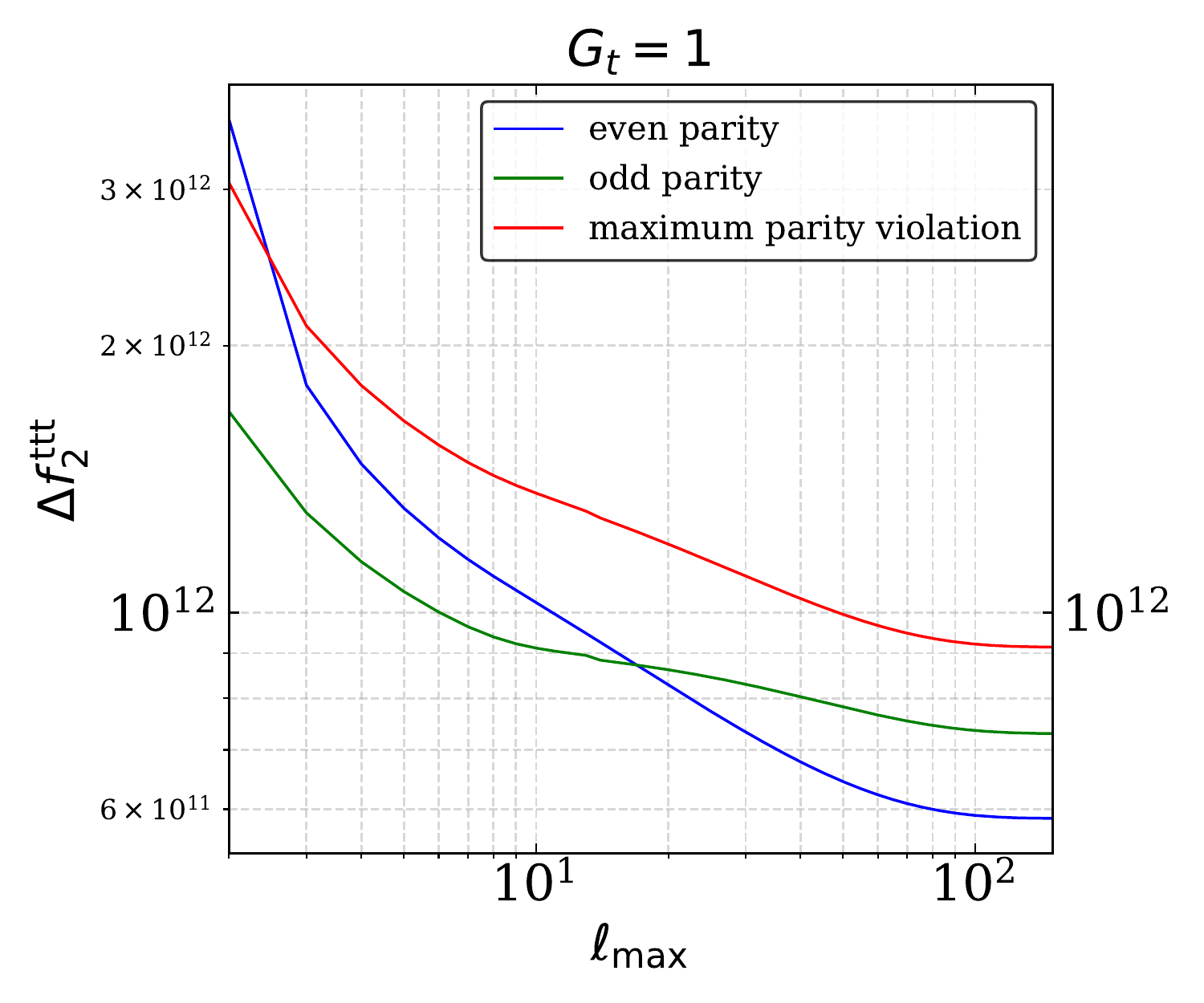} \,\, \includegraphics[width=.48\textwidth]{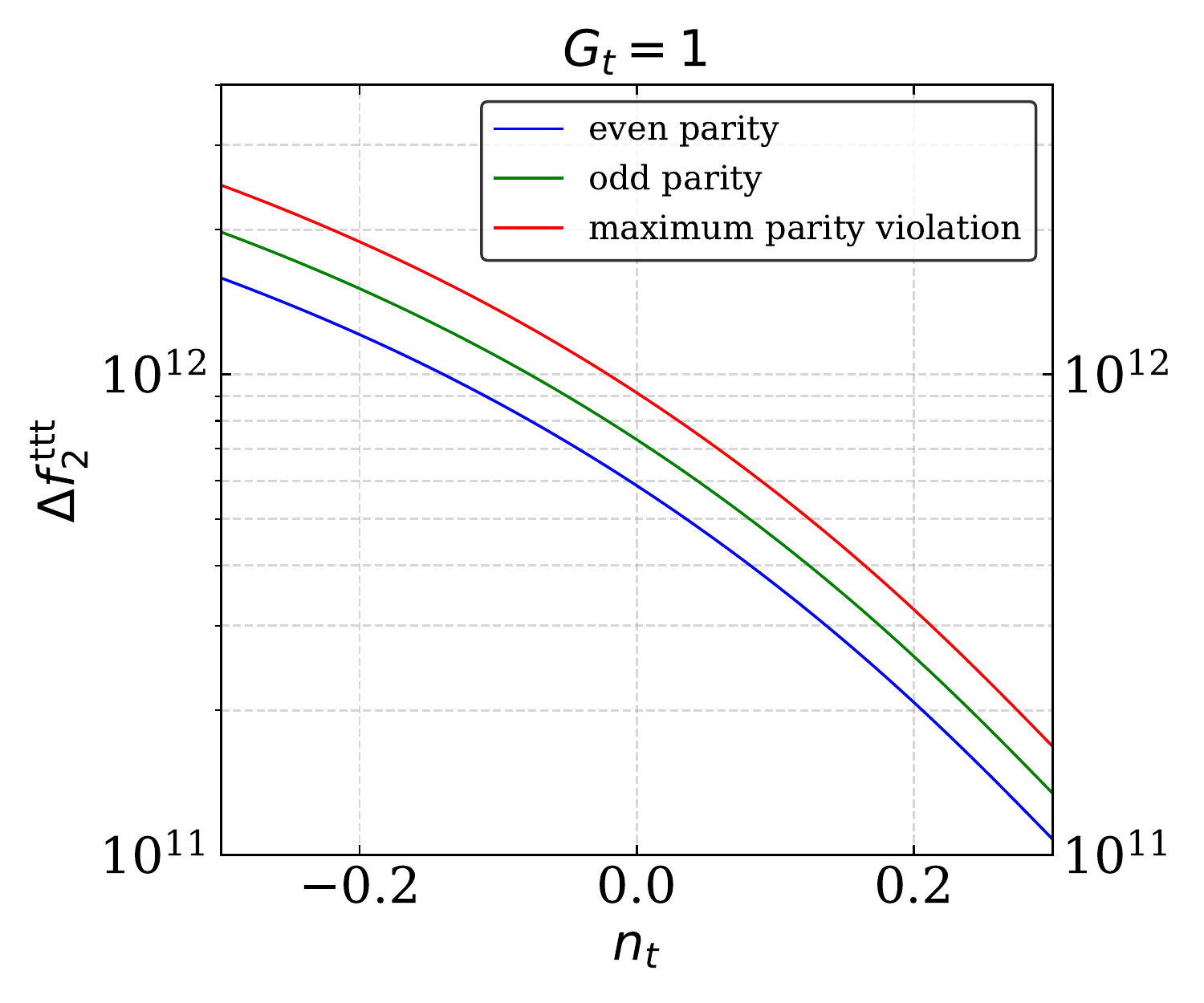}
  \caption{Plot of the expected 1-sigma error on $f^{\rm ttt}_2$ from $\mu B$ for a PIXIE-like level of noise and cosmic-variance limited fully delensed $B$ modes. The tensor-to-scalar ratio at CMB scales is taken $r_{\rm CMB} = 0.01$. Left panel: $\Delta f_2^{\rm ttt}$ vs $\ell_{\rm max}$ for scale invariant tensor power spectrum. Right panel: $\Delta f_{2}^{\rm ttt}$ vs $n_t$.
  \label{fig:Fisherttt}}
\end{figure}

Analogously to the previous subsection, the SD-CMB cross-correlation most sensitive to the 3-tensors primordial bispectrum is the $C^{\mu B}_{\ell_1 \ell_2 m_1 m_2}$ cross-spectrum. The Fisher matrix reads
\be
F = \sum_{\ell_1, \ell_2=2}^{\ell_{\rm max}}  \frac{\tilde A_{\ell_1 \ell_2}^{L M, \,\mu B}\,\tilde A_{\ell_1 \ell_2}^{*L M, \,\mu B}}{C^{\mu \mu}_{\ell_1, \rm N} \, C^{BB}_{\ell_2} } \, ,
\ee
where $\tilde A_{\ell_1 \ell_2}^{L M, \,\mu B} =  \partial A_{\ell_1 \ell_2}^{L M, \,\mu B}/\partial f^{ttt}_{L}$ and
\ba
A_{\ell_1 \ell_2}^{L M, \,\mu B} = &  \,\, i^{\ell_1-\ell_2} \,\sqrt{(2 \ell_1 + 1) (2 \ell_2 + 1)} \, \begin{pmatrix}
	\ell_1 & \ell_2 & L \\
	0 & 2 & -2
	\end{pmatrix} \nonumber \\
& \times \, 8 \pi \, \mathcal I^{\ell_1 \ell_2, L M, B}_{\gamma \gamma \gamma} \, ,
\ea
where $\mathcal I^{\ell_1 \ell_2, L M, B}_{\gamma \gamma \gamma}$ is as in eq. \eqref{IB:ttt}. Here we adopt the convention $|\xi_{RRR}|=|\xi_{LLL}| = 1$, and we assume $\xi_{RRL}= \xi_{LLR} = 0$ \footnote{As before, we are not including the contribution of squeezed bispectra with mixed chiralities. This can lead to an improvement up to a factor 2 of the 1-sigma error on $f_2^{ttt}$ in the even parity and odd parity cases of fig. \ref{fig:Fisherttt}.}. In fig. \ref{fig:Fisherttt} we plot the expected 1-sigma error on $f^{ttt}_{L}$ with $L = 2$ for the various ways in which the $\langle \gamma \gamma \gamma \rangle$ bispectrum transforms under parity transformation. 
From our plots and the Fisher matrix we obtain the following scaling formula
\ba \label{eq:Delta_ttt}
\Delta f_{2}^{ttt}|_{B} = \frac{a}{G_t} \, \left(\frac{\mu_{\rm N}}{4.96  \times 10^{-8}}\right) \, \left(\frac{0.01}{r_{\rm CMB}}\right)^{1/2} \, 10^{b(n_t) + c(n_t)^2} \, .
\ea
In tab. \ref{tab:fit_power_ttt} we summarize the values of the fit parameters $a$, $b$, $c$. 
\begin{table}[h!]
\begin{center}
\begin{tabular}{ |c|c|c|c| } 
\hline 
Transformation under parity  & $a$ & $b$ & $c$ \\ 
\hline
even parity & $5.9 \times 10^{11}$ & $- 1.91$ & $-1.52$ \\ 
\hline
odd parity & $7.3 \times 10^{11}$ & $- 1.91$ & $-1.52$  \\ 
\hline
maximum parity violation & $9.0 \times 10^{11}$ &  $- 1.91$ & $-1.52$ \\
\hline
\end{tabular}
\end{center}
\caption{Values of the parameters in eq. \eqref{eq:Delta_ttt}.} \label{tab:fit_power_ttt}
\end{table}
As before, we have a slight dependence on the level of parity violation, while the dependence on the other relevant parameters is not altered by the level of parity violation.

In the next section, we make some considerations about what we learn from these forecasts and their validity. We also discuss various classes of inflationary models that could be of relevance for the observables we are considering. 

\section{Model considerations}
\label{sec:observation}

In this section, we consider various phenomenological and model building aspects of our findings. We begin by commenting on the assumption of weak NGs. In performing our Fisher matrix forecasts in the previous section, we implicitly assumed that primordial NGs are small compared to the Gaussian part of the primordial correlators. Demanding that non-linear effects from a given squeezed primordial correlation $\langle x x' x'' \rangle$ are subdominant than the linear results, translates into the following condition
\be \label{eq:non_lin_cons}
k_l^3 k_s^3 \, \langle x(k_l) x'(k_s) x''(k_s)\rangle \ll \Delta_x(k_l) \Delta_{x'}(k_s) \Delta_{x''}(k_s) \, ,
\ee
where $\Delta_{i} = (\mathcal A_{i})^{1/2}$ denote the square roots of the dimensionless power spectra. By applying eq. \eqref{eq:non_lin_cons} to primordial bispectra we get the following constraints to our non-Gaussian amplitudes
\ba \label{eq:tts_limit_small_NG}
f_{L}^{tts}, \, f_{L}^{sss} &\ll  \Big(\mathcal A_{s}(k_{\rm CMB}) \Big)^{-1} \simeq  10^{5} \, , \\
f_{L}^{sst}, \, f_{L}^{ttt} &\ll \Big(\mathcal A_{s}(k_{\rm CMB}) \, r_{\rm CMB} \Big)^{-1} \simeq   \frac{10^{5}}{r_{\rm CMB}} \, .
\ea
These theoretical upper bounds should be matched with the 1-sigma error derived from the plots above. Assuming almost scale invariant spectra in the window of scales where primordial perturbations source $\mu$-distortions, we got the following rough expressions for a PIXIE-like experiment
\be \label{eq:Delta_scalings}
 \Delta f_{L}^{tts} \approx \frac{10^{10}}{G_T}  \left(\frac{0.01}{r_{\rm CMB}}\right)^{1/2}\, , \quad \Delta f_{L}^{sss} \approx \frac{10^{3}}{G_s}  \, , \quad \Delta f_{L}^{sst} \approx  \frac{10^{4}}{G_s} \left(\frac{0.01}{r_{\rm CMB}}\right)^{1/2} \, , \quad \Delta f_{L}^{ttt} \approx  \frac{10^{11}}{G_T} \left(\frac{0.01}{r_{\rm CMB}}\right)^{1/2} \, .
\ee
Assuming $r_{\rm CMB} \sim 0.01 -0.001$ and the absence of growth-mechanisms ($G_i = 1$), we get that the expected 1-sigma error on $\langle \gamma \gamma \gamma \rangle$ and $\langle \gamma \gamma \zeta \rangle$ is larger than the theoretical upper bound. This is something that was expected if we look at fig. \ref{fig:transfer_mu}. The tensor $\mu$ modes transfer function is about five orders of magnitude smaller than the scalar transfer function. Therefore, in absence of mechanisms of amplifications of primordial perturbations we would expect 
\be \label{eq:Delta_scalings2}
 \Delta f_{L}^{tts} \approx 10^{-5} \, r_{\rm CMB} \, \Delta f_{L}^{sss} \, , \qquad \qquad \Delta f_{L}^{ttt} \approx   10^{-5} \, r_{\rm CMB} \, \Delta f_{L}^{sst} \, ,
\ee
which is in agreement with \eqref{eq:Delta_scalings} for $G_i = 1$. From eq. \eqref{eq:Delta_scalings2} it follows that we need the following amplification of the tensor perturbations amplitude
\be 
G_T \approx \frac{10^{5}}{r_{\rm CMB}} 
\ee
in order for $\langle \gamma \gamma \gamma \rangle$ and $\langle \gamma \gamma \zeta \rangle$ bispectra to reach the same level of detectability as $\langle \zeta \zeta \zeta \rangle$ and $\langle \zeta \zeta \gamma \rangle$. In absence of such an amplification of tensor perturbations,  $\langle \gamma \gamma \gamma \rangle$ and $\langle \gamma \gamma \zeta \rangle$ bispectra are basically unconstrained by the cross-correlators between CMB $\mu$ modes and CMB temperature and polarization anisotropies. Given the current upper bound on $r_{\rm CMB}$ from the Planck experiment, we need an amplification factor of at least $G_T \gtrsim 10^6$, (and proportionally more if $G_s$ is also greater than one). Such a huge, independent amplification of tensor inflationary perturbations in the $\mu$ modes-era is typically not reachable within a controlled approximation in inflationary models known to us. As noticed e.g. in \cite{Kite:2020uix}, only gravitational waves of post-inflationary origin appear to be realistic targets for $\mu$-distortions experiments. This suggests that from our current vantage point $\langle \zeta \zeta \zeta \rangle$ and $\langle \zeta \zeta \gamma \rangle$ are most likely the only bispectra we may put realistic constraints using the cross-correlations we are considering. However, model builders may some day concoct a model that successfully independently amplifies primordial tensor perturbations with $G_T \gtrsim 10^6$.

By including the sky-damping factor $f_{\rm sky}$, we also get the scaling formulas 
\ba \label{eq:Delta_scalings_final}
&\Delta f_{L}^{sss}|_{T+E} \sim \frac{10^{3}}{G_s \, f_{\rm sky}} \left(\frac{\mu_{\rm N}}{4.96 \times 10^{-8}}\right)  \, 10^{-2.55(n_s-1) - 0.58(n_s-1)^2}  \, ,\nonumber \\
&\Delta f_{L}^{sst}|_{B} \sim  \frac{10^{4} }{G_s \, f_{\rm sky}} \left(\frac{\mu_{\rm N}}{4.96 \times 10^{-8}}\right) \left(\frac{0.01}{r_{\rm CMB}}\right)^{1/2} \, 10^{-2.55(n_s-1) - 0.58(n_s-1)^2}  \, .
\ea
Combining these equations we get the following scaling \footnote{We note that this scaling is model dependent only insofar as the spectra can be parametrized as a power law with a fixed (or weakly running index). There are a large class of models for which this isn't the case, necessitating a separate, though straightforward generalization of the present treatment.}
\be \label{eq:Delta_scalings_independent}
\Delta f_{L}^{sst}|_{B}  \sim  10 \left(\frac{0.01}{r_{\rm CMB}}\right)^{1/2}  \Delta f_{L}^{sss}|_{T+E}  \, .
\ee
Assuming $r_{\rm CMB} =0.01$ we get that $\mathcal O(1)$ squeezed non-Gaussian $\langle \zeta \zeta \zeta \rangle$ ($\langle \zeta \zeta \gamma \rangle$) amplitudes can be measured if the amplification of scalar perturbations satisfies $G_s > 10^3 \,(10^4)$. These results are particularly interesting since models with such amplification in the power spectra typically also share an analogous amplification of the non-Gaussian amplitudes. 

Examples of such scenarios are inflationary models of primordial black hole (PBH) production (see, e.g., \cite{Garcia-Bellido:2017mdw, Ezquiaga:2017fvi, Ballesteros:2017fsr,Hertzberg:2017dkh, Cicoli:2018asa,Ozsoy:2018flq,Mahbub:2019uhl, Ballesteros:2020qam}). As shown e.g. in \cite{Ozsoy:2021qrg}, though an ultra-slow-roll mechanism we can enhance the power spectrum of scalar perturbations up to seven orders of magnitude, reaching $G_s = 10^7$. 
Assuming such an amplification mechanism and an experiment with a PIXIE-like level of noise with $f_{\rm sky} = 0.1$ and $r_{\rm CMB} =0.01$, this would lead to $\Delta f_{L}^{sss} \simeq  10^{-3}$ and $\Delta f_{L}^{sst} \simeq  10^{-2}$. However, it is worth to stress that the presence of even a low level of statistical anisotropies in these bispectra is necessary to get non-trivial signatures. To our knowledge, the effects of statistical anisotropies in such models is still unexplored. 

Similarly, interesting detection prospects have already been considered in literature for the pure scalar (isotropic) bispectrum $\langle \zeta \zeta \zeta \rangle$ from the $\mu T$ and $\mu E$ cross-correlations (see e.g.~\cite{Pajer:2012vz,Ravenni:2017lgw,Ozsoy:2021qrg}). Here we want to emphasize that in those models in which statistical anisotropies in the $\langle \zeta \zeta \zeta \rangle$ bispectrum leave detectable signatures in $\mu T$ and $\mu E$, there is also a possibility to detect a non-zero $\langle \zeta \zeta \gamma \rangle$ bispectrum signal from the $\mu B$ cross-correlator. This is relevant as the $\langle \zeta \zeta \gamma \rangle$ bispectrum reveals information on the underlying inflationary scenario that are usually not contained in the $\langle \zeta \zeta \zeta \rangle$ bispectrum. For example, a measurement or constraint on $\langle \zeta \zeta \gamma \rangle$ would allow us to probe: (i) the interactions between scalar and tensor primordial perturbations, (ii) the gravitational waves induced by second order scalar perturbations, (iii) a deeper insight in the violation of rotational and parity symmetries in the primordial universe. In particular, this last feature is rather interesting. As we have shown in sec. \ref{sec:comments}, depending on the kind of statistical anisotropy and the way in which the $\langle \zeta \zeta \gamma \rangle$ bispectrum transforms under parity symmetry we are able to predict the multipole configurations that provide a non-zero signal. This implies that a detection of a non-zero $\langle \mu_{\ell_1} B_{\ell_2} \rangle$ signal in certain $\ell_1, \ell_2$ doublets may provide detailed information about the violation of the parity symmetry in inflationary models. Needless to say, a similar argument applies also to $\langle \mu_{\ell_1} T_{\ell_2} \rangle$ and $\langle \mu_{\ell_1} E_{\ell_2} \rangle$ cross-correlations. In this case, parity violation signatures may be left imprinted also by the $\langle \zeta \zeta \zeta \rangle$ bispectrum with $L_1 = \mbox{odd}$, $L_2 = 0$ statistical anisotropies. However, when $L_1 = \mbox{even}$, we must rely solely on $\langle \gamma \zeta \zeta \rangle$ to probe parity violation.   

In order for tensor NGs to be meaningfully detectable via SD-CMB cross correlations, one evidently requires a large amplification of scalar and/or tensor modes at scales relevant for $\mu$-distortions, which moreover, must be sourced by a background that also violates statistical isotropy. Although this may seem like a doubly contrived demand, there is evidently a class of models for which the violation of statistical isotropy and the amplification of primordial perturbations may go hand in hand. Inflation realized via a scalar field charged under a U(1) symmetry with an inflaton dependent gauge kinetic coupling has been studied by the authors of \cite{Emami:2009vd, Emami:2015qjl} as a means to generate observable levels of statistical anisotropy. The model action is given by
\be \label{eq:anisotropic_model}
S=\int d^{4} x \sqrt{-g}\left[\frac{M_{P}^{2}}{2} R-\frac{1}{2} D_{\mu} \phi D^{\mu} \bar{\phi}-\frac{f^{2}(\phi)}{4} F_{\mu \nu} F^{\mu \nu}-V(\phi, \bar{\phi})\right].
\ee
A non-zero expectation value for the gauge potential (which breaks isotropy) is sustained during inflation through a combination of the gauge kinetic mixing and the minimal coupling of the charged scalar to the U(1) field. The inflaton potential can correspond to a range of universality classes, including hilltop, hybrid, and chaotic inflation, implying a large degree of parametric freedom in this class of models \cite{Emami:2009vd, Emami:2015qjl}. The presence of higher dimensional terms (in the power counting sense) implicit in the operator $f^2(\phi)F^2$ forces us to consider the above as an effective action, for which the additional interaction $g^2(\phi)F\widetilde F$ appears with the same degree of (ir)relevance. The presence of the latter term, which for the modulus of the charged scalar mimics that of an axionic coupling to $F\widetilde F$ has been shown to generically source large, secondarily produced primordial perturbations (e.g. \cite{Cook:2013xea,Barnaby:2010vf}). Although such an iteration of the class of models represented by Eq.~\eqref{eq:anisotropic_model} has not been studied in the literature to our knowledge, it is of equal relevance from a power counting perspective, and places the violation of statistical isotropy and the generation of enhanced scalars and tensors on an equal footing. The parametric freedom in the choices of the three independent functions $f(\phi), g(\phi)$ and $V(\phi)$ ought to be suggestive to any interested model builders as a possibility to generate the level of SD-CMB cross correlations relevant to the considerations of this paper.

\begin{figure}[h!]  
\centering
   \includegraphics[width=.6\textwidth]{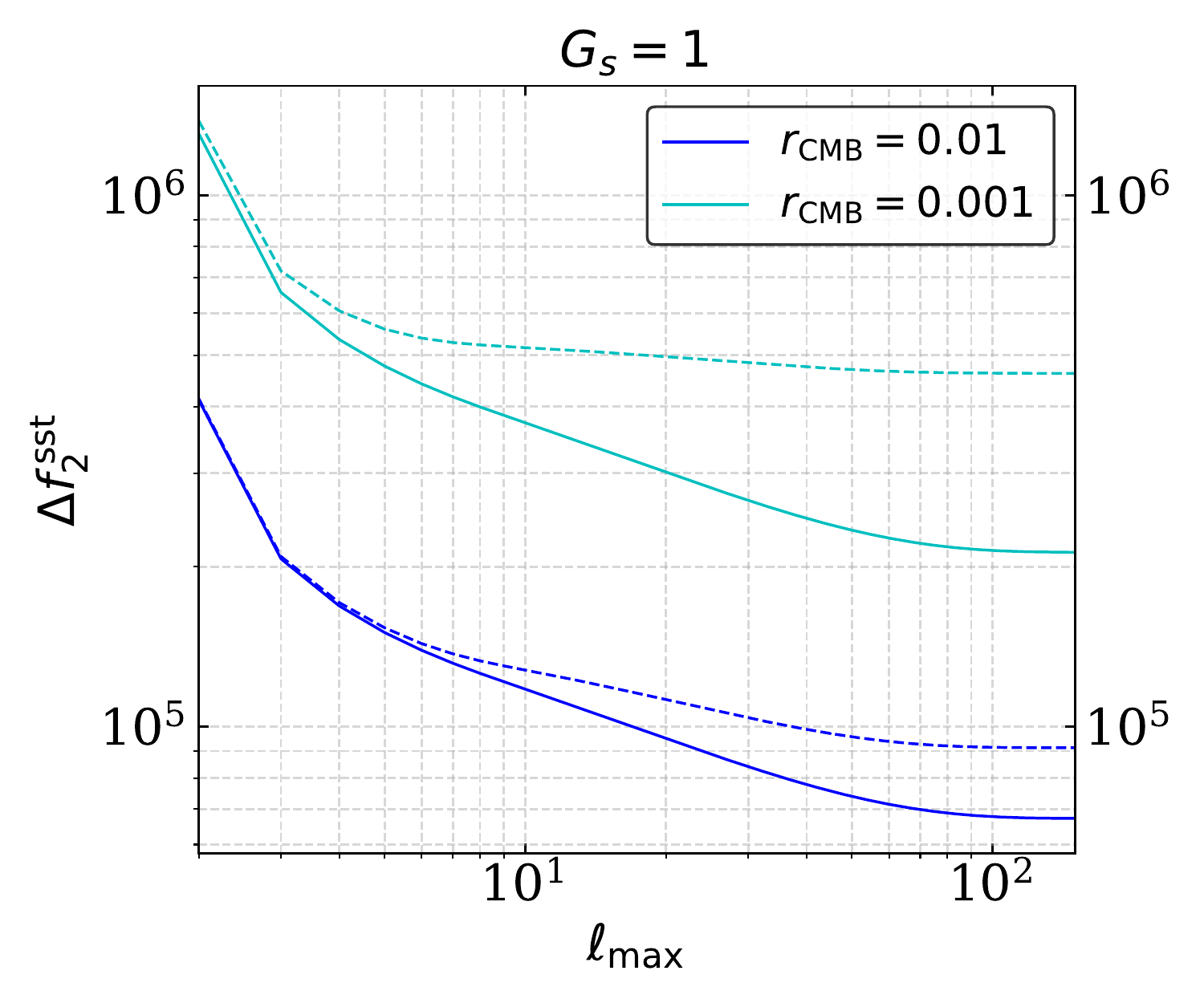} 
  \caption{Plot of the expected 1-sigma error on $f^{\rm sst}_2$ for the parity even case as in fig. \ref{fig:Fishersst} obtained for different values of the tensor-to-scalar ratio at CMB scales. Straight lines: fully delensed $B$ modes. Dashed lines: lensed $B$ modes. 
  \label{fig:Fisherlensing}}
\end{figure}

We conclude this section by discussing some limitations of our forecasts. First, we stress that the $\Delta f^{xxx}_{L}$'s derived above represent only the lower bound on the 1-sigma error, and they corresponds to the exact error only when our observables are Gaussian distributed. In our case the BipoSH coefficients $A_{\ell_1 \ell_2}^{L M, \mathcal O^1 \mathcal O^2}$ approach a Gaussian distribution only in the large-$\ell_1, \ell_2$ limit (see e.g. \cite{Joshi:2011vc}). Since we are looking to relatively large scale effects (low CMB multipoles), it is most likely that the real 1-sigma error is higher than what stated. Also, we did not account for the contribution of galactic foregrounds. As noticed in \cite{Abitbol:2017vwa,Remazeilles:2018kqd}, these should be taken into account for a real world experiment. On the other end, our forecasts are valid assuming that the non-Gaussian amplitudes $f^{xxx}_{L}$ are almost scale invariant functions of the parameters of an underlying inflationary model. As shown in eq. \eqref{eq:ansatzsss}, the squeezed limit amplitudes $f^{xxx}_{L}$ may depend on the short and soft modes $k_s$ and $k_l$. A scale dependence over $k_s$ and $k_l$ stronger than a logarithmic or a soft power law may modify our forecasts in a non-trivial way. In such a case one should reabsorb the scale dependence in a parameter $\alpha$ as
\be
f^{xxx}_{L}(k_s, k_l) = \alpha(k_s, k_l) \, \tilde f^{xxx}_{L} \, ,
\ee
where $\tilde f^{xxx}_{L}$ is a scale-invariant quantity. Therefore, we can reabsorb the quantity $\alpha(k_s, k_l)$ inside the momenta integrations of e.g. eq. \eqref{IT:sss}. The final forecast should be made on $\tilde f^{xxx}_{L}$. Finally, we like to discuss the degradation of the detectability of $\langle \zeta \zeta \gamma \rangle$ due to lensing contamination on CMB $B$ modes. By reproducing the plot in fig. \ref{fig:Fishersst} accounting for the contribution of the lensed $B$ modes in the cosmic variance limit we get the results summarized in fig. \ref{fig:Fisherlensing} \footnote{As an example, we only show the parity even case. We verified that the same qualitative conclusions arise when considering the other cases.}. As shown, for values of the tensor-to-scalar-ratio within the aim of the forthcoming CMB experiments ($r_{\rm CMB} = 0.01, 0.001$) the 1-sigma error increases at most a factor $2$, remaining of the same order of magnitude as the fully delensed case. This suggests that accounting for the lensing contamination in the cosmic variance limit will not significantly change the detectability prospects for the values of $r_{\rm CMB}$ that next generation of CMB experiments aims to measure.    

\newpage

\section{Conclusion} \label{sec:conclusion}

In this work we have explored new observational channels to probe tensor primordial NGs by exploiting the cross-correlations between the CMB $\mu$-distortions and temperature and polarization anisotropies. We have focused to the case where we introduce statistical anisotropies in squeezed NGs, as isotropic NGs leave either vanishing or highly suppressed signatures on the observables considered.

In detail, we have computed the effect of all the primordial (squeezed) bipectra involving both scalar and tensors on $\langle X_{\ell_1} \, \mu_{\ell_2} \rangle$ ($X = T, E, B$) cross-correlations. As statistical anisotropies in squeezed bispectra induce statistical anisotropies in these cross-correlations, we introduced the BipoSH formalism and BipoSH coefficients to study the detectability prospects through Fisher-matrix forecasts. 

We found that $\langle \zeta \zeta \gamma \rangle$ and $\langle \zeta \zeta \zeta \rangle$ are the only bispectra where we could realistically observe statistical anisotropies by cross-correlating the observational channels of the current and forthcoming CMB experiments, like {\it Planck}, LiteBIRD, PICO and PIXIE-like or proble class missions. The signatures left by the $\langle \gamma \gamma \gamma \rangle$ and $\langle \gamma \gamma \zeta \rangle$ bispectra are limited by the corresponding $\langle \zeta \zeta \gamma \rangle$ and $\langle \zeta \zeta \zeta \rangle$ bispectra unless a mechanism generates huge, independent growth in tensor perturbations in the window of scales where primordial perturbations source $\mu$-distortions. In particular, tensor perturbations have to increase by over six orders of magnitude with respect to the current constraints placed by the {\it Planck} experiment. What makes the detection of $\langle \gamma \gamma \gamma \rangle$ and $\langle \gamma \gamma \zeta \rangle$ very challenging is that tensor perturbations dissipate their energy much more inefficiently than scalar perturbations. 

We also found that for those inflationary models where the detection prospects of statistical anisotropies in the $\langle \zeta \zeta \zeta \rangle$ bispectrum are enhanced by combining the $\langle \mu_{\ell_1} T_{\ell_2}  \rangle$ and $\langle \mu_{\ell_1} E_{\ell_2}  \rangle$ information, also the detection prospects for probing statistical anisotropies in $\langle \zeta \zeta \gamma \rangle$ bispectrum from the cross-correlation $\langle \mu_{\ell_1} B_{\ell_2} \,  \rangle$ are enhanced (see the  model independent eq. \eqref{eq:Delta_scalings_independent}). This is relevant as it provides us with an observational channel complementary to the usual $\langle B B\rangle$ channel to find evidence of primordial gravitational waves in non-conventional models of inflation. 

Our final eq. \eqref{eq:Delta_scalings_final} aims to predict the level of detectability of the $\langle \zeta \zeta \zeta \rangle$ and $\langle \zeta \zeta \gamma \rangle$ non-Gaussian amplitudes defined in \eqref{eq:ansatzsss} and \eqref{eq:ansatzsst}.
Our forecasts are valid provided that in the window of scales where $\mu$ modes are produced we can approximate the scale dependence of the primordial power spectra as power-laws with almost constant spectral indexes, and assuming that the non-Gaussian amplitudes are almost constant. In case of more general scale-dependencies, a more detailed model-dependent analysis should be performed, using and adapting the results derived in sec. \ref{sec:comp_non_gaus}. However, the general claim that the detectability prospects of statistical anisotropies in $\langle \zeta \zeta \gamma \rangle$ bispectrum are enhanced wherever we realize a model able to enhance the detection prospects of statistical anisotropies in $\langle \zeta \zeta \zeta \rangle$ bispectrum, is universally valid, independently of the specific realization of inflation. 

As a final remark, we stress that a detection of a net signature on the cross-correlations we have considered in this work would imply a realization of inflation containing very peculiar features, like huge growth mechanisms of primordial perturbations in a very localized window of scales, and the presence of a controlled level of statistical anisotropies which must be consistent with the bounds already placed by the current CMB experiments. These are stringent constraints on inflationary model building, where any candidate model has to simultaneously realize these two conditions. Henceforth, a detection of a signature through these channels with forthcoming CMB experiments would most likely be sourced by non-standard inflationary dynamics, possibly within the class of models represented in Eq. \eqref{eq:anisotropic_model} with an additional interaction of the form $g^2(\phi)F\widetilde F$, although such a model has yet to be studied in the literature to our knowledge. We argue that the observational channels of primordial tensor NGs we proposed in this work will become important in a very futuristic scenario, where we will be able to exploit the cosmic variance limit level of noise in CMB experiments. In this regards, an analysis of the detection prospects of the signatures considered here in an ultimate survey is left for future research.

\paragraph{Acknowledgements}

We want to thank Giovanni Cabass, Jens Chluba, Ema Dimastrogiovanni, Enrico Pajer, Andrea Ravenni, Maresuke Shiraishi, and Gianmassimo Tasinato for useful comments on the draft. We are grateful to Enrico Pajer for constructive criticism on our preliminary results. G.O. and P.D.M acknowledge support from the Netherlands organization for scientific research (NWO) VIDI grant (dossier 639.042.730).
\newpage

\appendix

\section{Spin-raising and lowering operators and spin-weighted spherical harmonics} \label{appen:spin_operators}

Here, we briefly review the definitions of the spin-raising and lowering operators, giving an example on how we can use them to define the weighted spherical harmonics. We refer to e.g.~\cite{Zaldarriaga:1996xe} for more details. The spin raising $\up$ and lowering $\down$ operators acting on a generic spin s function ${}_sf(\theta,\phi) $ defined on a 2D sphere are given by
\begin{align}
\up {}_sf(\theta,\phi) &= -\sin^{s} \theta
\left[\partial_\theta + i\csc \theta
\partial_\phi \right]\sin^{-s} \theta
{}_sf(\theta,\phi)~, \nonumber\\
%---
\down {}_sf(\theta,\phi) &= -\sin^{-s} \theta 
\left[ \partial_\theta - i\csc \theta
\partial_\phi \right]\sin^{s} \theta {}_sf(\theta,\phi)~.
\label{eq:edth}
\end{align}
In particular, the new functions $\up {}_sf(\theta,\phi)$ and $\down {}_sf(\theta,\phi)$ have spin $s+1$ and $s-1$, respectively. For example, the spin raising and lowering operators acting twice on a generic
spin-$\pm 2$ function ${}_{\pm 2}f(\mu,\phi)$ which is factorized as ${}_{\pm 2}f(\theta,\phi) = {}_{\pm 2}\tilde{f}(\mu) e^{i m \phi}$ (i.e. the CMB polarization
fields) can be expressed as
\begin{align}
\down^2 {}_2f(\theta,\phi)&=
\left(-\partial_\mu + \f{m}{1-\mu^2}\right)^2 \left[
(1-\mu^2) {}_2f(\mu,\phi)\right] ~, \nonumber \\ 
%--- 
\up^2 {}_{-2}f(\theta,\phi) &=
\left(-\partial_\mu - \f{m}{1-\mu^2}\right)^2 \left[(1-\mu^2) {}_{-2}
f(\mu,\phi)\right]~, 
\label{eq:operators1}
\end{align}
where $\mu \equiv \cos \theta$. In this way, just acting with a differential operator, we can easily define spin-0 quantities starting from spin-2 ones. This procedure is used in the case of CMB to pass from the $P^{\pm}$ spin-$\pm2$ linear polarization fields to the $E$ and $B$ modes, which are spin-0 fields.

Using eqs. \eqref{eq:edth}, we can express the spin-weighted spherical harmonic functions on a 2D sphere, ${}_sY_{\ell m}(\theta,\phi)$, in terms of the common spherical harmonics ${}_{0}Y_{\ell m}(\theta, \phi) = Y_{\ell m}(\theta, \phi)$ by acting with the spin raising/lowering operator as 
\begin{align} \label{eq:spin_harm_sph}
{}_sY_{\ell m}(\theta,\phi) &=
\left[\f{(\ell-s)!}{(\ell+s)!}\right]^{\f{1}{2}}\up^s  Y_{\ell m}(\theta,\phi)
\ \  (0 \leq s \leq \ell) ~, \nonumber  \\ 
{}_sY_{\ell m}(\theta,\phi) &=
\left[\f{(\ell+s)!}{(\ell-s)!}\right]^{\f{1}{2}}(-1)^s 
\down^{-s} Y_{\ell m}(\theta,\phi)
\ \  (-\ell \leq s \leq 0) ~.
\end{align}
Then, it is possible to show the validity of the following relations
\begin{align}
\up {}_sY_{\ell m}(\theta,\phi) &=\left[(\ell-s)(\ell+s+1)\right]^{\f{1}{2}}
\,{}_{s+1}Y_{\ell m}(\theta,\phi) ~, \nonumber \\
%---
\down {}_sY_{\ell m}(\theta,\phi) &=-\left[(\ell+s)(\ell-s+1)\right]^{\f{1}{2}}
\,{}_{s-1}Y_{\ell m}(\theta,\phi) ~, \nonumber \\
%----
\down\up {}_sY_{\ell m}(\theta,\phi) &=-(\ell-s)(\ell+s+1)
\,{}_{s}Y_{\ell m}(\theta,\phi) ~m \, ,
\label{eq:propYs}
\end{align}
which can be used to derive the following explicit expression of the weighted spherical harmonics
\begin{eqnarray}
{}_sY_{\ell m}(\theta, \phi) &=& e^{im\phi}
\left[\f{(\ell+m)!(\ell-m)!}{(\ell+s)!(\ell-s)!}
\f{(2\ell+1)}{4\pi}\right]^{1/2}
\sin^{2\ell}(\theta/2) \nonumber \\
&&\times \sum_r {\ell-s \choose r}{\ell+s \choose r+s-m}
(-1)^{\ell-r-s+m}{\rm cot}^{2r+s-m}(\theta/2) ~.
\label{eq:expl}
\end{eqnarray} 
It is straightforward to verify the orthogonality and completeness conditions for the ${}_sY_{\ell m}(\theta, \phi)$ as 
\begin{align} \label{eq:orto_harm}
\int_0^{2\pi} d\phi \int_{-1}^1 d\cos \theta \,\, {}_s Y_{\ell' m'}^*(\theta,\phi)
\, {}_s Y_{\ell m}(\theta,\phi) &= \delta_{\ell', \ell} \, \delta_{m', m} ~, \nonumber \\
\sum_{\ell m} {}_s Y_{\ell m}^*(\theta,\phi) \,
{}_s Y_{\ell m}(\theta',\phi')
&= \delta(\phi-\phi')\delta(\cos\theta-\cos\theta')\, ,
\end{align} 
as well as the following properties regarding the transformation under conjugate and parity
\begin{align}
{}_sY^*_{\ell m}(\theta, \phi) &= (-1)^{s + m}{}_{-s}Y_{\ell -m}(\theta, \phi)~, \nonumber \\
%---
{}_s Y_{\ell m}(\pi - \theta, \phi + \pi) &= (-1)^{\ell} \, {}_{-s} Y_{\ell m}(\theta, \phi)~. 
\label{eq:parity_Y}\end{align}

\section{3-j symbols, Gaunt integral and Clebsch-Gordan coefficients} \label{app:Wigner}

In this appendix, we give some useful formulas regarding the angular integrals of products of spherical harmonics. We will use $\hat x$ to denote a given direction on the 2D sphere and $d^2 \Omega_x$ to indicate the infinitesimal solid angle on the sphere.

First, we define the quantity ${}_{s_1 s_2 s_3} \mathcal G_{\ell_1 \ell_2 \ell_3}^{m_1 m_2 m_3}$, which is known as "generalized" Gaunt integral and it represents the angular integral of the product of three (weighted) spherical harmonics. This can be written in terms of Wigner 3-j symbols as (see e.g. \cite{Komatsu:2003iq, Liguori:2005rj})
\begin{align} \label{eq:Gaunt_integral}
{}_{s_1 s_2 s_3}\mathcal G_{\ell_1 \ell_2 \ell_3}^{m_1 m_2 m_3} &= \int d^2 \Omega_x \, {}_{s_1}Y_{\ell_1 m_1}(\hat x) \, {}_{s_2}Y_{\ell_2 m_2}(\hat x)  \,{}_{s_3}Y_{\ell_3 m_3}(\hat x) \nonumber\\
&= \sqrt{\f{(2 \ell_1 + 1) (2 \ell_2 + 1) (2 \ell_3 + 1)}{4 \pi}}   \begin{pmatrix}
	\ell_1 & \ell_2 & \ell_3 \\
	-s_1 & -s_2 & -s_3
	\end{pmatrix}\begin{pmatrix}
	\ell_1 & \ell_2 & \ell_3 \\
	m_1 & m_2 & m_3
	\end{pmatrix} \, .
\end{align}
The Wigner 3-j symbols are related to the spin-weighted spherical harmonics as
\begin{align} \label{eq:rel_wigner}
\prod_{i = 1}^2 {}_{s_i}Y_{\ell_i m_i}(\hat x)&= \sum_{\ell_3 m_3 s_3} \,  {}_{s_3}Y^*_{\ell_3 m_3}(\hat x) \, \sqrt{\f{(2 \ell_1 + 1) (2 \ell_2 + 1) (2 \ell_3 + 1)}{4 \pi}}   \nonumber\\
& \times \begin{pmatrix}
	\ell_1 & \ell_2 & \ell_3 \\
	-s_1 & -s_2 & -s_3
	\end{pmatrix}\begin{pmatrix}
	\ell_1 & \ell_2 & \ell_3 \\
	m_1 & m_2 & m_3
	\end{pmatrix} \, .
\end{align}
Notice that eq. \eqref{eq:Gaunt_integral} follows once putting together eqs. \eqref{eq:orto_harm} and \eqref{eq:rel_wigner}. 

Some useful properties of the Wigner 3-j symbols that we used in this work are
\begin{align} \label{eq:lwigner}
 \begin{pmatrix}
	\ell_1 & \ell_2 & \ell_3 \\
	m_1 & m_2 & m_3
	\end{pmatrix} = (-1)^{\sum_i \ell_i}\begin{pmatrix}
	\ell_1 & \ell_2 & \ell_3 \\
	 - m_1 & - m_2 & - m_3
	\end{pmatrix} \, ,
\end{align}
and
\begin{align} \label{eq:sum_m_wigner}
\sum_{m_1, m_2}  \begin{pmatrix}
	\ell_1 & \ell_2 & \ell_3 \\
	m_1 & m_2 & m_3
	\end{pmatrix}\begin{pmatrix}
	\ell_1 & \ell_2 & \ell'_3 \\
	m_1 & m_2 & m'_3
	\end{pmatrix}  = (2 \ell_3 + 1)^{-1} \, \delta_{\ell_3, \ell'_3} \, \delta_{m_3, m'_3} \, .
\end{align}
The 3-j symbols of the kind
\be \label{eq:3jsymbols}
\begin{pmatrix}
	\ell_1 & \ell_2 & \ell_3 \\
	m_1 & m_2 & - m_3
	\end{pmatrix}
\ee
are related to the Clebsh-Gordan coefficients
\be \label{eq:CG}
\mathcal C^{\ell_2 m_3}_{\ell_1 m_1 \ell_2 m_2} = \langle \ell_1  m_1 \ell_2 m_2| \ell_3 m_3  \rangle
\ee
by \cite{Shiraishi:2012}
\be \label{eq:3jsymbols-CG}
\begin{pmatrix}
	\ell_1 & \ell_2 & \ell_3 \\
	m_1 & m_2 & - m_3
	\end{pmatrix} =  \frac{(-1)^{\ell_1 - \ell_2 + m_3}}{\sqrt{2 \ell_3 + 1}} \, \mathcal C^{\ell_2 m_3}_{\ell_1 m_1 \ell_2 m_2} \, .
\ee
Therefore, the 3-j symbols of the form \eqref{eq:3jsymbols} vanish unless the selection rules are satisfied as follows
\ba
&|m_1| \leq \ell_1 \,, \qquad  |m_2|\leq \ell_2 \,, \qquad |m_3|\leq \ell_3 \,, \qquad  m_1 +m_2 =m_3 \, ,\\
&|\ell_1 − \ell_2| \leq \ell_3 \leq \ell_1 + \ell_2 \quad \mbox{(the triangle condition)} \, , \qquad \ell_1 +\ell_2 + \ell_3 \in Z \, .
\ea
More properties of the Wigner 3-j symbols can be found in \cite{Shiraishi:2012}.

\bibliography{references.bib}
\end{document}